\pdfoutput=1
\documentclass[11pt,a4paper]{article}
\usepackage{array}
\usepackage[small,bf]{caption} 
\usepackage[dvips]{graphicx}
\usepackage{color} 
\usepackage{xcolor} 
\usepackage{colortbl}
\usepackage{amsmath}
\usepackage{fancyhdr}
\usepackage[]{natbib}
\usepackage[LGRgreek]{mathastext}
\usepackage[yyyymmdd,hhmmss]{datetime}
\usepackage{longtable}

\newcolumntype{L}[1]{>{\raggedright\let\newline\\\arraybackslash\hspace{0pt}}m{#1}}
\newcolumntype{C}[1]{>{\centering  \let\newline\\\arraybackslash\hspace{0pt}}m{#1}}
\newcolumntype{R}[1]{>{\raggedleft \let\newline\\\arraybackslash\hspace{0pt}}m{#1}}
\newcolumntype{T}[1]{>{\vbox to 0ex\bgroup\vfill\centering}p{#1}<{\egroup}}  

\fancyheadoffset[R]{0pt}
\fancyfootoffset[R]{0pt}

\definecolor{dred}  {rgb}{0.6,0.0,0.0}
\definecolor{aqgr}  {rgb}{0.0,1.0,0.6} 
\definecolor{viol}  {rgb}{0.8,0.6,0.8}
\definecolor{figdr} {rgb}{1.0,1.0,1.0} 
\definecolor{colnu} {rgb}{1.0,0.0,1.0} 
\definecolor{colhd} {rgb}{1.0,0.8,0.0} 

\newcolumntype{C}[1]{>{\centering\let\newline\\\arraybackslash\hspace{0pt}}m{#1}}
\renewcommand{\arraystretch}{1.8}
\title{\bfseries{\textsc{Towards a general model \\ for psychopathology}}}

\author{Alessandro Fontana} \date{}

\begin{document}
\maketitle
   
\clubpenalty=10000
\widowpenalty=10000

\begin{abstract}
The DSM-1 was published in 1952, contains 128 diagnostic categories, described in 132 pages. The DSM-5 appeared in 2013, contains 541 diagnostic categories, described in 947 pages. The field of psychology is characterised by a steady proliferation of diagnostic models and subcategories, that seems to be inspired by the principle of ``divide and inflate''. This approach is in contrast with experimental evidence, which suggests on one hand that traumas of various kind are often present in the anamnesis of patients and, on the other, that the gene variants implicated are shared across a wide range of diagnoses. In this work I propose a holistic approach, built with tools borrowed from the field of Artificial Intelligence.  
My model is based on two pillars. The first one is trauma, which represents the attack to the mind, is psychological in nature and has its origin in the environment. The second pillar is dissociation, which represents the mind defence in both physiological and pathological conditions, and incorporates all other defence mechanisms. Damages to dissociation can be considered as another category of attacks, that are neurobiological in nature and can be of genetic or environmental origin. They include, among other factors, synaptic over-pruning, abuse of drugs and inflammation. These factors concur to weaken the defence, represented by the neural networks that implement the dissociation mechanism in the brain.
The model is subsequently used to interpret five mental conditions: PTSD, complex PTSD, dissociative identity disorder, schizophrenia and bipolar disorder. Ideally, this is a first step towards building a model that aims to explain a wider range of psychopathological affections with a single theoretical framework. The last part is dedicated to sketching a new psychotherapy for psychological trauma.
\end{abstract}


\pagebreak

\tableofcontents

\newpage
\section{Introduction}  
\label{sec:introduction}

The DSM-1 (1952) contains 128 categories, described in 132 pages.
The DSM-2 (1968) contains 193 categories, described in 119 pages.
The DSM-3 (1987) contains 253 categories, described in 567 pages.
The DSM-4 (2000) contains 383 categories, described in 943 pages.
The DSM-5 (2013) contains 541 categories, described in 947 pages \citep{Blashfield2014}.

Some might argue that the proliferation of diagnostic categories is a sign of a growing comprehension of mental physiology and pathology. I am convinced that the opposite is true. The ultimate goal of scientific research is to explain phenomena with the least number of the simplest models (Occam's razor). The proliferation of diagnostic models and subcategories seems to be inspired by the principle ``divide and inflate'', with reference to the fact that the categories created by the last division have a tendency to inflate themselves, until another division is needed. This approach is in contrast with experimental evidence, that suggests on one hand that traumas of various kind are always present in mental disease and, on the other, that the underlying genetics is shared across a wide range of conditions.

I choose to approach the problem through the method of Artificial Life (``study life as it could be, to understand life as we know it'', \citet{langton1989}) and with the toolset of Artificial Intelligence. Artificial neural networks are a computational model that represents the state-of-the-art in many AI applications. A neural network is composed of a number of computational units called neurons, arranged in layers, that learns through examples with a method called ``deep learning''. Artificial neural networks draw inspiration from biological neural networks: in this work, we reverse the inspiration flow and apply deep learning to the source of its inspiration: the brain itself.

Following these ideas, I propose a general model that seeks to explain several mental conditions with a unified framework, based on two pillars. The first one is trauma, which represents the attack to the mind, is psychological in nature and has its origin in the environment. The second pillar is dissociation, which represents the mind defence in both physiological and pathological conditions, and incorporates all other defence mechanisms. Damages to dissociation can be considered as another category of attacks, that are neurobiological in nature and can be of genetic or environmental origin. They include, among other factors, synaptic over-pruning, abuse of drugs and inflammation. These factors concur to weaken the defence, represented by the neural networks that implement the dissociation mechanism in the brain. The model is subsequently used to interpret five mental conditions: PTSD, complex PTSD, dissociative identity disorder, schizophrenia and bipolar disorder.  

This paper is divided into six parts: this first section is the introduction; the second section outlines a possible model for brain architecture (this part can be skipped without compromising the understanding of the rest); the third section is dedicated to the model of the mind and its functioning in normal conditions; the fourth section introduces a reinterpretation of the concept of dissociation; the fifth section deals with post-traumatic disorders; the sixth section explores schizophrenia and bipolar disorder; the seventh section sketches the psychotherapy for traumas; the last section draws the conclusions and outlines future research directions. The mind is a very complex object, in both physiological and pathological manifestations. The model we are going to present is based on a simplification of reality, aimed at evidencing some limited aspects of mental functioning. 

\clearpage

\newpage
\section{Brain computational architecture}  

\subsection{Cortex and hippocampus}  

In the brain different types of memory are implemented. Based on duration, memory can be classified in short-term (or working) memory and long-term memory \citep{baddely2007}. Based on content, memory is classified in declarative (further subdivided in episodic and semantic) and implicit or procedural \citep{squire2009}. Regarding learning, the number of ways in which the brain can learn exceeds the power of any classification system.  

The hippocampus (a seahorse-shaped brain structure located in the medial temporal lobe) and the cortex (a 3 mm-thick layer of tissue distributed on the surface of the brain) are both involved in the process of memory formation. Overall, the empirical evidence seems to hint that the hippocampus stores complete and unprocessed memory records for a short time, while the cortex develops features capturing high-level generalisations of data \citep{preston2013}, that are stored for longer (possibly indefinite) periods. These generalisations may correspond to the schemas of Piaget's developmental theory \citep{piaget1952}. 

Hippocampi are located (more or less) in the centre of each brain hemisphere, in a region where fibres carrying multiple sensory inputs converge \citep{amaral1995}. Each hippocampus is directly connected to a portion of cortex called entorhinal cortex, which functions as a connection hub to and from the rest of cortex. The size of the entorhinal cortex is relatively small and (presumably) only allows a subset of cortical fibres to reach the hippocampus \citep{canto2008}.  

A decisive contribution to our understanding of memory processes came from studies conducted on patient H.M., who had large portions of both medial temporal lobes (including the hippocampi) removed at age 27, in an attempt to treat severe epilepsy. The surgical procedure was successful in treating epilepsy, but left the patient with an almost complete anterograde amnesia and a graded retrograde amnesia. In other words, H.M. was unable to form new declarative memories and his recollection of recent past events was impaired, while older memories were intact \citep{scoville1957}.   

Based on these and other studies, the ``standard model'' of memory formation and consolidation was proposed and consolidated \citep{squire1986, mcclelland1995}. The model foresees that new (declarative) memories are first stored in the hippocampus and then, in a process that can last decades, gradually transferred to the cortex, where they are stored indefinitely. Once the transfer is complete, memories are retained even if the hippocampus is removed or damaged. 

The vast majority of brain neurons are generated during embryonic development. However, new neurons are created also in specific regions of the adult brain. The generation of neurons in the dentate gyrus of the hippocampus is a process that continues for the whole duration of life \citep{eriksson1998}, favoured by physical exercise \citep{nokia2016}. It is natural to think that such neurons are involved in the process of memory formation in this brain region.  

A recent model \citep{kitamura2017} suggests that the instantiation of the cortical representation of memory ``engrams'' \citep{bruce2001} occurs from the very beginning. New memories, instead of being first recorded in the hippocampus and then gradually copied or moved to the cortex, would be written in both places in parallel. The cortical representation would be immature at first and develop with time to more mature forms.

\subsection{Artificial neural networks}  

Artificial neural networks (simply neural networks from now on) are computational models that take inspiration from biological neural networks. The ``multilayer perceptron'' \citep{rosenblatt1958} is a particular type of neural network composed of a number of computational units called neurons, grouped in layers stacked on top of each other. Input and output layers are always present, if the number of intermediate layers is greater than one, the network is called a deep neural network. Deep neural networks represent the state of the art in many AI applications, such as computer vision, speech recognition, natural language processing. The algorithm used for training is called ``backpropagation'' and the method ``deep learning''.

Neural networks learn from data, organised in a data set, structured as a collection of records. An example of data set (once) used to train neural networks is the Mnist \citep{lecun1998}, a collection of 60000 images representing hand-written characters from 0 to 9, where each record of the data set is a set of numbers (the data set variables) encoding the grey shades of the 28x28 individual pixels that compose an image. From the data set, the network learns higher-level features, representing oriented edges or simple shapes, combination of simple shapes, and the ``concepts'' of digits.        

\begin{figure*}[t] \begin{center} \hspace*{-0.50cm}
\includegraphics[width=18.00cm]{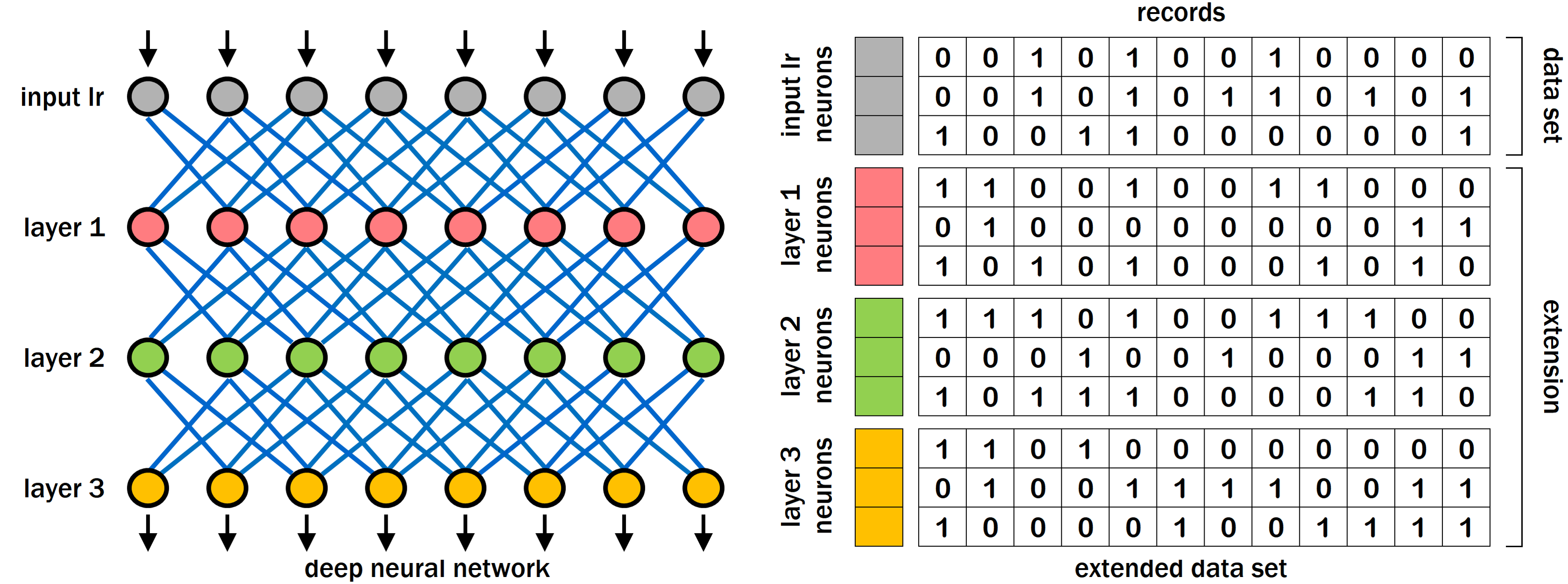}
\caption{Left: structure of a deep neural network. The network is composed of computational units called neurons, arranged in layers stacked on top of each other (input and output layers are always present, the number of intermediate layers defines the network's depth). Right: example of extended data set. Rows correspond to features /neurons, columns correspond to records. Input neurons encode the proper data set features, neurons of layer 1 and subsequent layers encode new features acquired by the neural network as a result of the learning process.}
\label{deepneds}
\end{center} \end{figure*}

For learning, data set variables are mapped to neurons of the neural network input layer, while high-level features are encoded in neurons of successive layers (Fig.~\ref{deepneds}, left). Learning consists in modifying the parameters that define the strength of connections between neurons which, collectively, represent the memory of the network. Neural networks can learn to perform a variety of tasks (e.g., recognition, classification, etc.): once learning is finished, such tasks can be performed on new data very quickly. 

The typical learning procedure is divided in two parts: training (in which connection parameters are optimised based on the data set) and test (in which the network is tested on data not used for training). The training part is in turn structured as a cycle composed of two phases: in the ``change'' phase the algorithm brings small changes to connection parameters, while in the ``assessment'' phase the network performance is measured on the data set, yielding a performance score. This two-phase cycle (called ``epoch'' in the machine learning jargon) is repeated a number of times, until the score reaches a satisfactory value. 

With ``backpropagation'' (the most commonly used supervised learning algorithm), the score depends on the percentage of records classified correctly for output neurons and on the ``propagated'' gradient of the classification error for neurons of intermediate layers; with unsupervised learning algorithms, other scores are used. While in supervised learning all neurons are optimised at the same time, in unsupervised learning, e.g. ``contrastive divergence'' \citep{hinton2002}, neurons are usually optimised one layer at a time: first layer 1 neurons, then layer 2 neurons, etc. 

Let us suppose that unsupervised learning is used and layer 1 neurons are being optimised: Fig.~\ref{deepneds}, right, shows all the data structures needed in the training process. The assessment phase procedure requires that: i) input neurons are exposed to all data records, ii) outputs of layer 1 neurons are calculated for all data records and iii) the score is calculated for the neurons that are being optimised. 

This requires that the entire data set is \textit{read} at each assessment phase during training (first row block in Fig.~\ref{deepneds}, right). Furthermore, computing the outputs of layer 1 neurons (needed to obtain the score) requires that these outputs are also \textit{written} (second row block in the figure). These additional rows are strictly-speaking not part of the data set, but correspond to the new features learned, whose values must be calculated for each record (each column). 

It is convenient to define the \textbf{extended data set} as the union, for each data record, of all values of data set variables and successive layers' neurons (Fig.~\ref{deepneds}, right). With this definition, we can conclude that the training procedure requires that the extended data set be accessed \textit{in both reading and writing} at each assessment phase during training. This, in turn, requires the storage of the extended data set for the entire duration of the process. 

Some training algorithms may be able to process data ``on the fly'', with each data record seen once and then discarded. In principle this is possible, but the most commonly used algorithms, proved to be effective in practical applications, need to be exposed to data many times. We will argue that this need is shared also by the ``algorithms'' operating in the brain.  

\subsection{Implementation of deep networks in the brain}  

\begin{figure*}[t] \begin{center} \hspace*{-0.0cm}
\includegraphics[width=17.00cm]{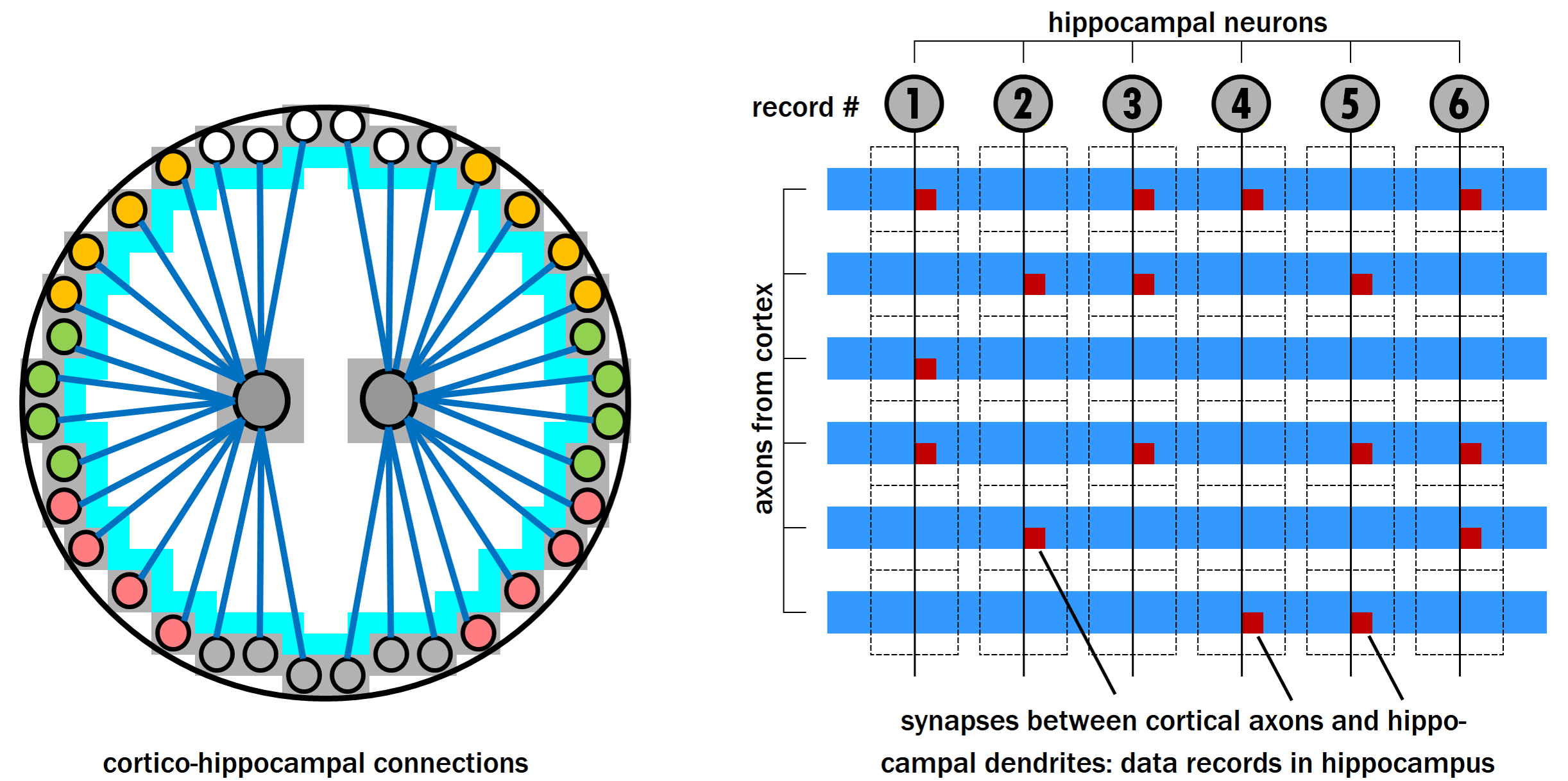}
\caption{Left: Proposed brain architecture. Features are implemented as neurons in the cortex. Neurons are organised in layers, shown with different colours: input (sensory) neurons in grey, layer 1 neurons in red, layer 2 neurons in green, layer 3 neurons in yellow, etc. Layer n neurons' inputs are connected with the outputs of neurons of all previous layers through lateral connections (shown as a light blue belt around the cortex). Neurons send also their axons to the hippocampi (blue radial links). Right: Creation of new associations among cortical neurons in the hippocampus. The axons of all cortical neurons, representing as many features, are projected to the hippocampus, where they are very close to each other (much closer than their respective cellular bodies). When a set of neurons fire together, their axons make connections to the dendrites of a hippocampal neuron (possibly a new neuron generated in the dentate gyrus): in this way, a new record of the ``brain data set'' is created.}
\label{ball}
\end{center} \end{figure*}

Understanding and interpreting reality is the main objective of brain activity, and reality can be conceptualised in terms of \textbf{features}, defined as descriptors or qualities that can be used to characterise a given situation. Features can be simple visual qualities, such as ``red colour'', ``round shape'', ``vertical orientation'', or correspond to more complex visual-motor characteristics, such as ``dance movements'', ``hiding an object with the hand'', etc. The model of reality is constructed by establishing meaningful associations among features: such associations become features in their own right, and can take part in further associations.

The term ``feature'' is borrowed from the field of artificial neural networks \citep{cox2014} where, in the case of e.g. visual applications, simple features may represent basic visual characteristics (oriented edges), complex features represent parts of objects (eyes, nose, mouth, etc.), more complex features represent whole objects (faces), or scenes consisting of many objects. In this context, the term refers also to ideas and abstract thoughts: in practice, features are the components of mental life in its broadest sense.

Our hypothesis is that features of any complexity are implemented in the brain by single neurons \citep{fontana2017a}. A neuron may represent a basic visual or auditory feature or the most complex philosophical thought: the complex ``meaning'' of a feature is given by its relation with all other features. We further conjecture that an association among features A, B and C is physically built by recruiting a new neuron N and linking through synapses the axons of neurons A, B and C to the dendrites of N, which encodes a new feature representing the association.    

The assumption of local coding (one-to-one mapping between features and single neurons in the brain) is consistent with experiments that show how neurons can exhibit selective response to very complex features, such as the face of a famous actress \citep{quiroga2005}. There is also evidence of whole neuronal populations involved in decision making and other complex behaviours \citep{david2017}. Current empirical evidence is insufficient to conclude in favour of or against the local coding hypothesis, which is commonly used in computational modelling (including neural networks). 

Neurons encoding features are located in the cortex (Fig.~\ref{ball}, left). Visual features, for instance, are mapped to the occipital cortex, auditory features are mapped to the auditory cortex, tactile features are mapped to the parietal cortex, etc. Neurons are connected with each other through ``lateral'' connections (light blue belt around the cortex in the figure) and are arranged in layers. Here the term ``lateral'' is used to distinguish such connections, which run tangentially along the cortex, from the radial ones discussed later: however, these connections implement the topological structure of Fig.~\ref{deepneds}, left.  

The neuronal organisation is not ``strictly layered'' as for the network in Fig.~\ref{deepneds}, left, in which layer n neurons' inputs are only connected with layer n-1 neurons' outputs. In this case layer n neurons' inputs are connected with the outputs of neurons of all previous layers, down to sensory input neurons. Also recurrent connections, that are known to play a role in the brain, are allowed in our model, which aims to be as general as possible. We assume that the ``cabling'' linking all cortical neurons is laid out during embryonic development, before learning starts.  

The axons of all cortical neurons project also to the hippocampus, where their tips are very close to each other, much closer than their respective cellular bodies (also these connections are pre-set before the start of the learning process). When a set of neurons fire together, their axons make synaptic connections to the dendrites of a hippocampal neuron generated in the dentate gyrus (Fig.~\ref{ball}, right): in this way, a new record of the ``brain extended data set'', encoding the co-occurrence of cortical activations, is created. 

We may also hypothesise that, at each moment in the course of life, neurons can be divided in two categories: already optimised neurons and neurons that are being trained. In the beginning, the first category is composed of sensory neurons only, while layer 1 neurons are being trained (in the case of visual learning, for example, the hippocampal data set is initially written by sensory neurons representing pixel intensity values, while the concepts of digits are learned). We may also hypothesise that neurons, which long ceased to be optimised, are gradually ``decommissioned'', until their axons lose access to entorhinal cortex and hippocampus (in the case of visual learning, once the concepts of digits become available, neurons encoding individual retinal intensity values would cease to be used to create new associations, and possibly be physically disconnected).

As already suggested in \citep{squire1986}, having all neurons project their axons to a small region is the only method to realise a fast association between features encoded in neurons that can be dispersed across a vast (in brain terms) cortical area. If the neurons are distributed on the surface of a sphere (as it is in the brain case), the best solution is that their axons project to a central point. In this point, the tips of all axons are very close to each other and the establishment of connections between them can be done very quickly. And it \textit{must} be done quickly, to keep up with the rapid flow of information fed from the perceptual apparatus. 

Based on these considerations, we suggest that the function of the hippocampus is to register and store the brain extended data set, structured as an array of records composed of features, encoded in cortical neurons, that co-occurred. From these records, the ``algorithms'' of the cortex extract high-level features, encoded in other cortical neurons through adjustments of their lateral connections. Based on the two-phase algorithmic structure described in section 2, we assume that the optimisation procedure requires the continuous access to the hippocampal formation by cortical neurons, in both reading and writing. The generation of new features and the addition of new records to the hippocampus are continuous processes that run in parallel. Cortical memories are instantiated from the beginning in immature form, and gradually develop to more mature forms, exactly as described in \citep{kitamura2017}. 


\subsection{Considerations on biological plausibility}  

Our model of the hippocampus as the extended data set of the brain is consistent with the finding that hippocampal neurons do more than encoding the spatial context \citep{eichenbaum2014} and goes beyond: we think that hippocampal neurons not only include space and time features, but all features. The model also accounts for the involvement of the hippocampus in flexible cognition and social behaviour \citep{rachael2014}, a task that requires the continuous updating of information and presupposes the continuous access to the hippocampus in both reading and writing.    

\citet{mcclelland1995} carry out a very thorough analysis of the functioning of biological memory compared to neural networks, concluding that ``interleaved learning'' is the most plausible model for the kind on learning that takes place in the brain. The authors argue that ``neural networks or connectionist models adhere to many aspects of the account of the mammalian memory system, but do not incorporate a special system for rapid acquisition of the contents of specific episodes and events''. 

Our counterargument to this observation is that neural networks do have a ``special system for rapid acquisition of the contents of specific episodes and events'' and this is nothing else than the computer memory section where the extended data set is stored. The motivation for the failed recognition of this component is probably the notion that data are strictly not part of the neural network model: they are however part of the system. 

In order to evaluate the biological plausibility of the model proposed, a quantitative assessment of the ``hardware'' requirements is needed. The model foresees that a hippocampal data record is implemented by a hippocampal neuron, and that each each activation of a cortical neuron corresponds to a connection between the hippocampal dendrite(s) and the cortical axon. Therefore, in order to accomodate a data set of N records in M variables, N hippocampal neurons are needed, each with M dendritic connections. If we hypothesise the acquisition of a new record every 10 seconds, we have (considering 12 hours of registration per day) a total of (1/10)*60*60*12*365 = 1576800 record per year, or 15768000 for ten years. According to \citet{west1990}, the dentate gyrus of a human hippocampus contains approximately $15*10^6$ neurons, able to support 10 years of registration based on the previous calculation. This seems to indicate that the model is indeed physically plausible.

The number of connections that need to be established for each record corresponds to the number of active features in that moment. This number is much smaller than the total number of features mapped in the brain (sparse coding). Furthermore, we can hypothesise the presence of additional elements to constrain and limit the number of active neurons per record. For instance, the mechanisms of attention and salience implemented in the limbic system \citep{bromberg2010} might filter the features that need to be recorded and reduce their number. In any case, according to our model, the number of active features per record only affects the number of dentritic connections for each hippocampal neuron, and not the number of hippocampal neurons that need to be recruited. Thus, even from this perspective, the model looks plausible.

Another potential issue is due to the fact that, in the real world, time is not discretised into bins and neurons in the brain behave asynchronously. It is not obvious how neuron activations occurring in partially overlapping intervals can be associated to the same record, which is the time scale of the records (a record every 1, 10, 20 seconds?), and whether the biological properties of neural plasticity are compatible with the hypothesised data throughput. This is indeed not only an issue for the proposed model, but for the learning process itself, related to what has been called ``distal reward problem'' or ``credit assignment problem'' \citep{sutton1998}.   

The most straightforward evidence for the biological plausibility of the asynchronous acquisition of episodic memory is the fact that it does happen, even if the neural details of the process are not fully elucidated. We remember who was present at the party, which food was served and which music was played, despite the fact that these features were not presented in the same ``time slot'', and cleanly and unambiguously associated to the same record. Several mechanisms can be proposed to explain this phenomenon: the features could be recorded in the same record, or they could be recorded in adjacent records, leaving the formation of associations or cause-effect relations to subsequent processing steps. A model for learning of associations between asynchronous stimuli is outlined in \citep{soltoggio2014}.    

H.M. had both hippocampal regions removed and, as a result, was unable to form new declarative memory: however, he was able to acquire new procedural memory. Different kinds of tasks fall under the definition of procedural memory: repetition priming, classical conditioning (Pavlov's experiments), emotional conditioning, various skills and habits such as mirror tracing, mirror reading or jigsaw puzzles. H.M.'s performance was good in many of these tasks \citep{woodruff1993}.

An explanation for the fact that procedural learning can take place without hippocampus is that procedural tasks might involve other brain regions (e.g., sensory cortex, cerebellum, amygdala, striatum). Furthermore, the presence of a stored data set might not always be required. Learning could still be possible by processing each new record (or a small number or records, depending on the storage capacity available without hippocampus) ``on the fly'', changing the neurons' parameters and then testing the performance of the modified neurons on subsequent records. This would explain the ability of amnesiacs to learn often-repeated material gradually over time \citep{corkin2002, glisky1986, mcclelland1995}.

\clearpage \pagebreak[4]

\clearpage

\newpage
\section{A model for the mind}  
\label{sec:modelmind}

\subsection{Cognitive features and social value}

Reality can be conceived as a set of situations, each characterised through a set of active \textbf{features}. Examples of simple perceptual features are: ``pyramidal shape'', ``vertical orientation'', ``colour green''. Examples of more abstract features are: ``blue swans'', ``baseball players'', ``being a plumber''. Feature activation and deactivation is a continuous process, driven by perceptual features fed from sensory stimuli and propagated to more abstract ones in real time. This occurs on a fast time scale, as the mind ``navigates'' through everyday life.

\begin{figure}[h] \begin{center} \hspace*{-0.00cm}
{\fboxrule=0.0mm\fboxsep=0mm\fbox{\includegraphics[width=15.00cm]{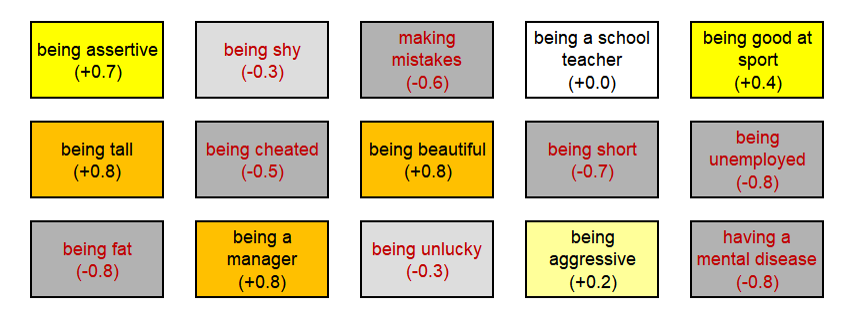}}}
\caption{Feature values. Our mind is full of ideas or features used to describe the world. Features have an associated property called social value, that can be positive or negative (to different degrees).}
\label{featvalues}
\end{center} \end{figure}

We postulate that features are characterised by a property called \textbf{social value}, that can be positive, e.g. ``honesty'' and ``health'', or negative, e.g. ``deception'' and ``illness'' (Fig.~\ref{featvalues}). The fact that features can be perceived as either positive or negative is known to all: the origin of value is less obvious. A clue can come from the observation of the behaviour of social animals, who pay much attention to the role of hierarchy: the organisation of packs of lions, wolves and many other species is based on the shared knowledge of the rank of each member. Human beings are no exception: in many situations, it is essential to know or decide who is superior and who is inferior. One hypothesis is that the mind is provided with an innate mechanism that allows to understand, in a social context, which are the group leaders and which are the followers. To features associated to the first ones, the mind attributes a positive value, to features associated to the second ones, a negative (or less positive) value.

Regardless of its origin, for most features value attribution seems to be intrinsically relative. This can be appreciated by considering the diversity of beliefs and reference values across ethnic groups, cultures, countries and historical periods. The word ``senate'', for instance, comes from the latin word ``senex'', which means ``old''. Nowadays, being old is a very disgraceful condition, but the fact that the ancient Romans called their most important institution ``house of the old'' is a hint that, in their society, being old was not so shameful. Another example is the idea ``being tanned'', which in the western world has a positive connotation. In the past, it was quite the opposite: paleness was a desirable quality. This because poor farmers, who had to work in the open, were naturally tanned, while upper class members used to walk around in golf courts with nice umbrellas to protect themselves from the sun. We owe the present (positive) value of the feature ``being tanned'' to Coco Chanel. 

A high diversity in values can be observed also in the same historical period and in the same country, across different social networks. In a given family, the feature ``being an artist'' may be considered ``cool'' and appreciated, while ``being an engineer'' may be considered boring and worthless. For a different family, the opposite may be true. The value background of a person is initially set by the senior family members (usually the parents), but can change over time.   

The change of values can happen by association: if a feature is associated with positive (negative) features, its value will become more positive (negative), not unlike what happens when two objects of different temperature are put in contact. This principle is used by marketing and advertising, when a product (of neutral value) is presented by (and thus associated with) a celebrity (having a very positive value). Since a feature in general takes part in many associations stored in the memory of everyone of us, its value will be determined by the combined effect of all associations. Value is a long term property, expected to change on a slow time scale. 

\subsection{Social emotions}

Many theories on the role and origin of emotions exist \citep{izard2009}, as well as computational models of the process of emotion generation \citep{marsella2010}. We hypothesise that emotions were evolved to serve three basic needs: survival (task involving one individual), reproduction (task involving two individuals) and social functioning (task involving many individuals). The first two needs are a direct translation of the requirements of Darwinian evolution and are shared by most living beings. The third derives from the necessity to organise a population of individuals and is also very common in the animal kingdom, probably because social organisation increases the Darwinian success of the species. 

\textbf{Fear} is probably the most primitive emotion and serves the first need: avoiding physical damage and escaping death. The expression of fear does not require a relational or social context: we can have fear even if we are alone in the world. Fear (or anxiety) plays also another special role among emotions: it serves to avoid potentially harmful events (fear of dying), but also any other ``bad'' emotion: fear of shame, fear of guilt, fear of pain, etc. (this is the feeling experienced by some during, e.g., public speaking).

\textbf{Love, jealousy and romantic pain} (the pain caused by the loss of a romantic /sexual partner, by either abandonment or death) serve the second need: finding and keeping a partner for reproduction. The expression of such emotions makes sense only in a relational context, but does not require a social one: as portrayed in numerous movies, we can feel love even if we are stranded with our partner on a deserted island. 

Social emotions serve the third need: preserving the organisation of a society of individuals. These emotions are often produced as a reaction to another person's actions in a social context and depend on the assessment of who is superior and who is inferior, who is right and who is wrong. Our hypothesis is that social emotions are synthetic social /hierarchical judgements, and that the determination of social value constitutes a precondition for the generation of such emotions.

\begin{figure}[t] \begin{center} \hspace*{-0.50cm}
{\fboxrule=0.0mm\fboxsep=0mm\fbox{\includegraphics[width=18.00cm]{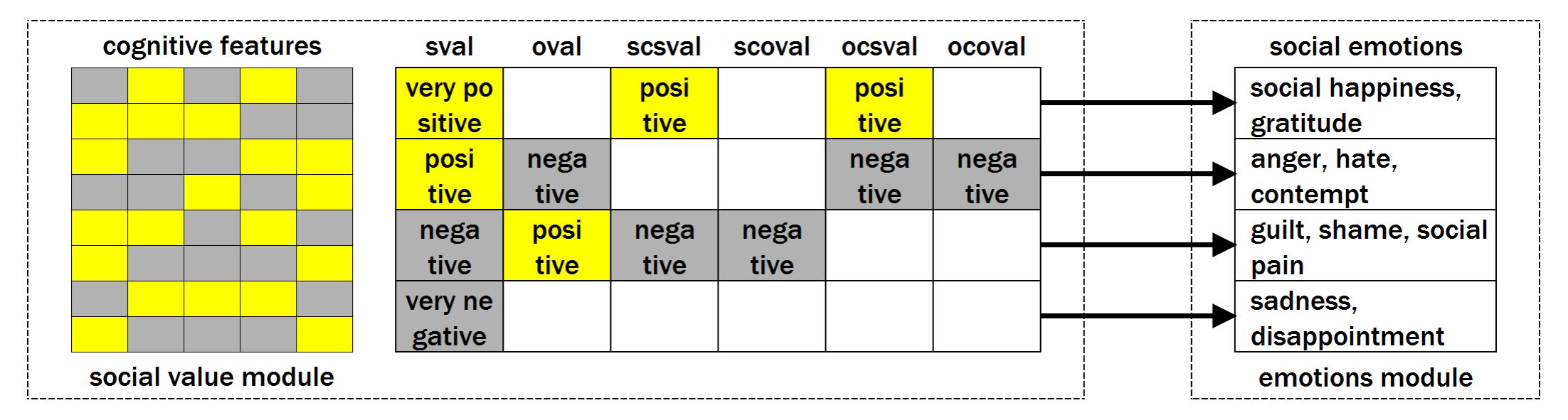}}}
\caption{Social value and emotions. Social emotions are based the computation of social value, a task performed by the social value module, which attributes value to features and computes the value of self (sval), object (oval), and four other auxiliary variables (scsval, scoval, ocsval, ocoval). Based on the computation results, the emotional module produces a specific emotion.}
\label{emotpoles}
\end{center} \end{figure}

More specifically, we foresee the existence of a module in the human mind dedicated to the computation of social value (Fig.~\ref{emotpoles}, left). This module assigns value to active features and calculates \textbf{sval} (self value) and \textbf{oval} (object value), and four other auxiliary variables: \textbf{scsval} (self contribution to self value), \textbf{scoval} (self contribution to object value), \textbf{ocsval} (object contribution to self value), \textbf{ocoval} (object contribution to object value). 


Self value and object value are in turn determined by the sum of the values of all \textit{active} features associated to them. Since the set of these features changes depending on the situation, so does the self value (as well as the object value). Therefore, in our model the self value (or self-esteem) is a spatio-temporally local concept: one person can have a positive self value in a given situation and a negative self value in another one. We can still define a global self value as the mean self value across all situations. If the self is stably associated to a core set of features, the mean value of such features determines what we might call the person's \textbf{mood}, which can also be positive or negative.

Based on these variables' values, the emotional module (Fig.~\ref{emotpoles}, right) generates an emotion. \textbf{Happiness}, for example, is generated when sval (self value) is very positive. \textbf{Gratitude} is generated when sval is very positive and ocsval is positive (in other words: when the self value is positive thanks to an action done by the object). \textbf{Anger} is produced when sval is positive, oval is negative and ocsval is positive (in other words: when the object's action causes a decrease in self value).

\textbf{Shame} occurs when sval is negative, oval is positive and scsval is negative (in other words: when the self has a value lower than the object and performs an action that reduces its own value). A positive object value is a precondition for the elicitation of shame: we don't feel ashamed if our mistakes are witnessed by a person who has a low hierarchical value (e.g. a small child). \textbf{Guilt} is generated when sval is negative, oval is positive and scoval is negative (in other words: if our actions cause a decrease in object value). \textbf{Sadness} is generated when sval (self value) is very negative.

So far it has been assumed that emotions depend on self and object values. We could ask if emotions depend not only on value, but also on value \emph{change}, as our intuition might suggest: disappointment seems e.g. to arise when something bad happens, causing a change (a decrease) in self value. Our answer is that the value changes associated to the emotion are in fact due to the change of situation, and emotions only depend on values (which, we recall, are spatio-temporally local).

\begin{figure}[t] \begin{center} \hspace*{-0.00cm}
{\fboxrule=0.0mm\fboxsep=0mm\fbox{\includegraphics[width=17.00cm]{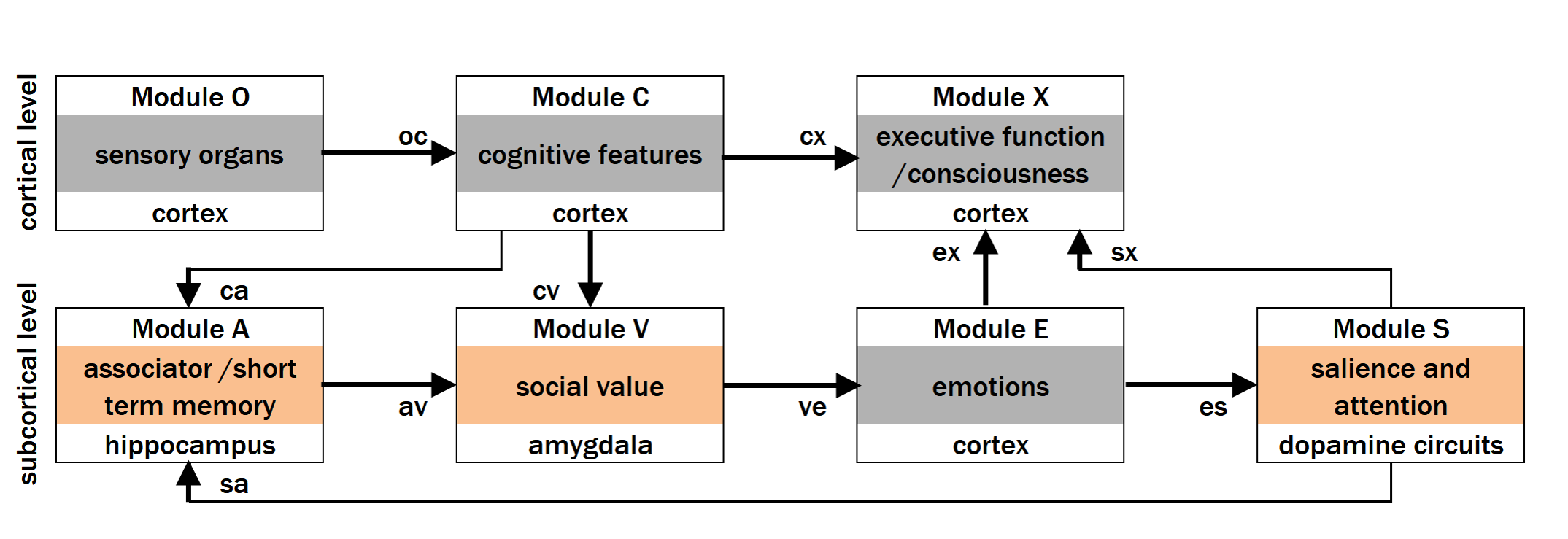}}}
\caption{Process of emotion generation. Cognitive features are activated and sent to consciousness and in parallel to the associator, where their coactivations are recorded. The next processing step is implemented by the social value module, where cognitive features are loaded with value. Based on this information, the emotions module produces an emotion that is presented to consciousness. Emotions generate salience, which is also presented to consciousness.}
\label{modules}
\end{center} \end{figure}

We assign social emotions to four groups, referred to as to ``poles'': A (emotions characterised by positive self value), B (emotions characterised by negative self value and positive object value), C (emotions characterised by negative self value) and D (emotions characterised by positive self value and negative object value). The association of shame, guilt and sadness to a lower social status is confirmed by numerous studies \citep{stevens1996}, as well as the association of anger to a higher social status \citep{tiedens2011}. As we shall see, only emotions belonging to groups B and C (in which the self value if negative) can be involved in a trauma.

The author is aware that the proposed scheme is not complete and is not able, in its current form, to make sense of the complexity of human emotions. Its main purpose is to convey the idea that social emotions are the outcome of a computational process involving the social value of self and others, rooted in the evolutionary need of social animals to establish a hierarchy among group members. Taking into account these limitations, we think it can still be useful to provide an intuition about the origin and purpose of emotions.   

\subsection{Mind modules, neurobiological correlates}  

The process of emotion generation can be represented through a set of interacting modules (Fig.~\ref{modules}). Cognitive features are activated by the sensory input and sent to consciousness and to the associator, where their coactivations are recorded. The next processing step is implemented by the social value module, where cognitive features are loaded with value. Based on this information, the emotions module produces an emotion that is presented to consciousness. Emotions generate salience, which is also presented to consciousness.

The scheme proposed contains the implicit assumption that memory records are exclusively composed of cognitive features. Emotions are not recorded in memory, but generated ``on-the-fly'' based on the set of active features, once such features are activated, through either external input or memory recall. This explains how the emotions associated to an event can change with time (e.g., painful memories become less painful as a result of therapy) even though the cognitive trace of the event persists (although even cognitive memories, both traumatic and non-traumatic, can change over time).

Based on current neurobiological knowledge, the processing of perceptual and abstract features occurs in the cortex (in e.g. occipital cortex for visual features), while the executive function is implemented in the prefrontal cortex. The associator corresponds to the hippocampus, a structure essential for the formation of new memories. Social values could be encoded in the amygdala which, according to recent studies \citep{bzdok2013}, would encode ``good'' and ``bad'' signals and would be indispensable for understanding social hierarchy. The emotional module could be located in the striatum, the salience module could be implemented by dopamine circuits.

\begin{figure}[t] \begin{center} \hspace*{-0.50cm}
{\fboxrule=0.0mm\fboxsep=0mm\fbox{\includegraphics[width=18.00cm]{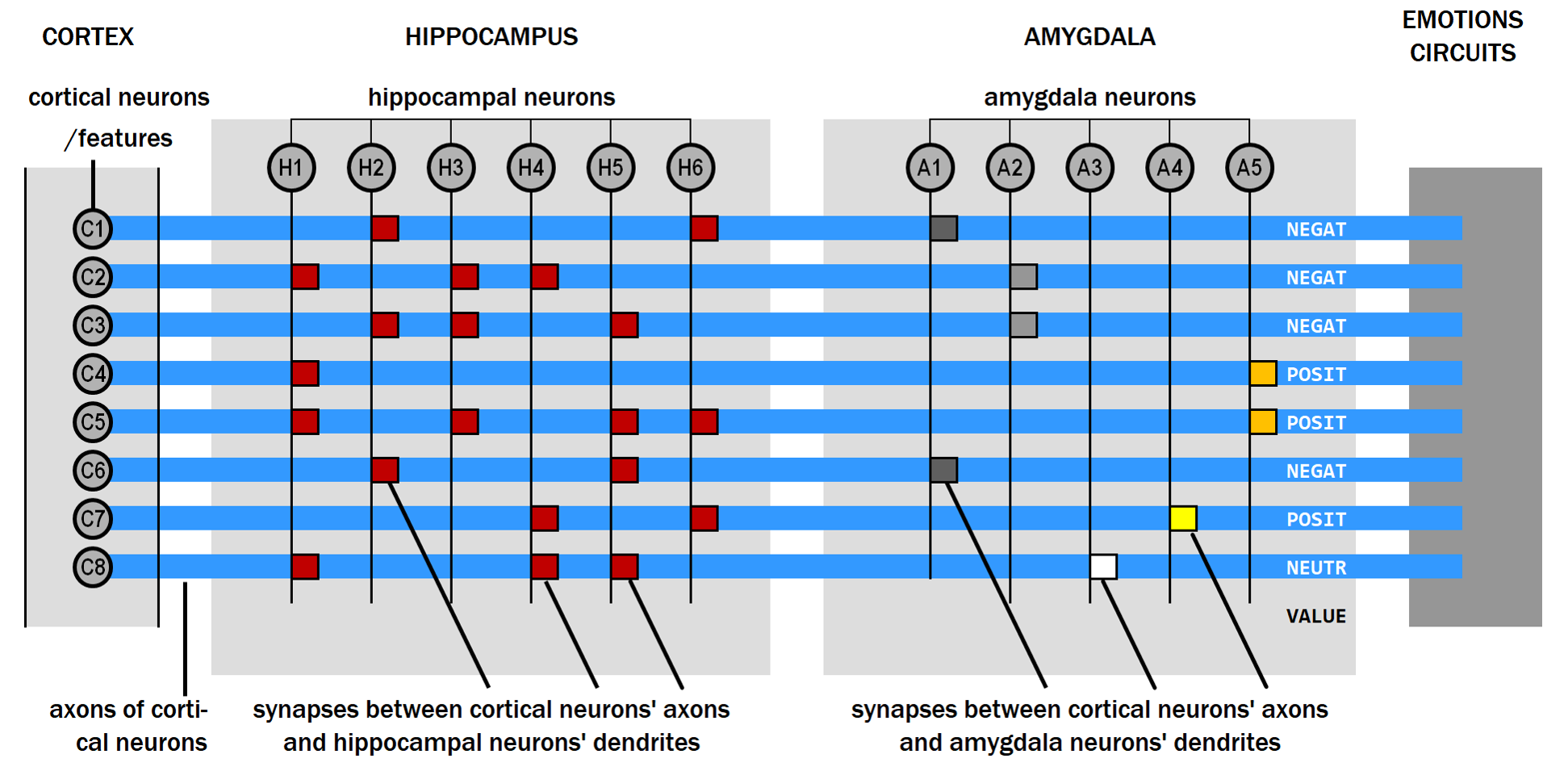}}}
\caption{Cortical neurons representing cognitive features send their axons to the hippocampus, where their co-activations are recorded in the synaptic weights between cortical neurons' axons and hippocampal neurons' dendrites. After leaving the hippocampus, the axons reach the amygdala, where the social values of features are encoded in the synaptic weights between cortical neurons' axons and amygdala neurons' dendrites. Based on the encoded values, emotions are produced by specific circuits.}
\label{hippamigd}
\end{center} \end{figure}

Fig.~\ref{hippamigd} shows a possible physical implementation of associator and social module in hippocampus and amygdala respectively. We hypothesise that cognitive features are implemented by single neurons in the cortex (e.g. occipital cortex neurons for visual features, temporal cortex neurons for auditory features, etc.). This assumption is consistent with experiments that show how neurons exhibit selective response to very complex features, such as the face of a famous actress \citep{quiroga2005}. Cortical neurons send their axons to the hippocampus, through the entorhinal cortex acting as a ``hub'' \citep{canto2008}. After leaving the hippocampus, the path of these axons continues into the amygdala (located near the hippocampus).

In the hippocampus, the co-activations (associations) of features are recorded and stored in the synaptic weights between cortical neurons' axons and hippocampal neurons' dendrites. Each set of co-activated features constitutes a record of the ``brain data set'', from which new, more complex features can be learned and encoded in other cortical neurons. In the amygdala, the social values of features are encoded in the synaptic weights between cortical neurons' axons and amygdala neurons' dendrites. 

In our model, the emotional module is not plastic. In other words, given the social value of active features, the emotional response is fixed. The plasticity of the emotional response resides in the associator and in the social module: synapses in the hippocampus are shaped by the input flow, and synapses in the amygdala are determined by the combined effect of the associations recorded in the hippocampus. If, for example, negative feature C1 co-occurs in hippocampal record H6 with positive features C5 and C7, the positive value propagates from C5 and C7 to C1 (and the negative value propagates from C1 to C5 and C7). The current values correspond to the point of equilibrium determined by the combined effect of all such ``value flows''.

\clearpage

\newpage
\section{Dissociation}  
\label{sec:Dissociation}

\subsection{Trauma and dissociation}  

A \textbf{situation} is defined as a set of features simultaneously active. Examples of situations are: ``piano lesson with uncle'' (coactive features: image of piano, image of hands on the keyboard, sound of uncle's voice, sound of piano, etc.); ``tennis match with a friend'' (coactive features: image of racquet, image of opponent, sound of racquet hitting the ball, odour of sweat, etc.). Situations can be thought of as belonging to different \textbf{contexts}, such as ``parent relation'', ``romantic relation'', ``relation with schoolmates'', etc..

\begin{figure*}[h] \begin{center} \hspace*{-0.00cm}
{\fboxrule=0.0mm\fboxsep=0mm\fbox{\includegraphics[width=16.50cm]{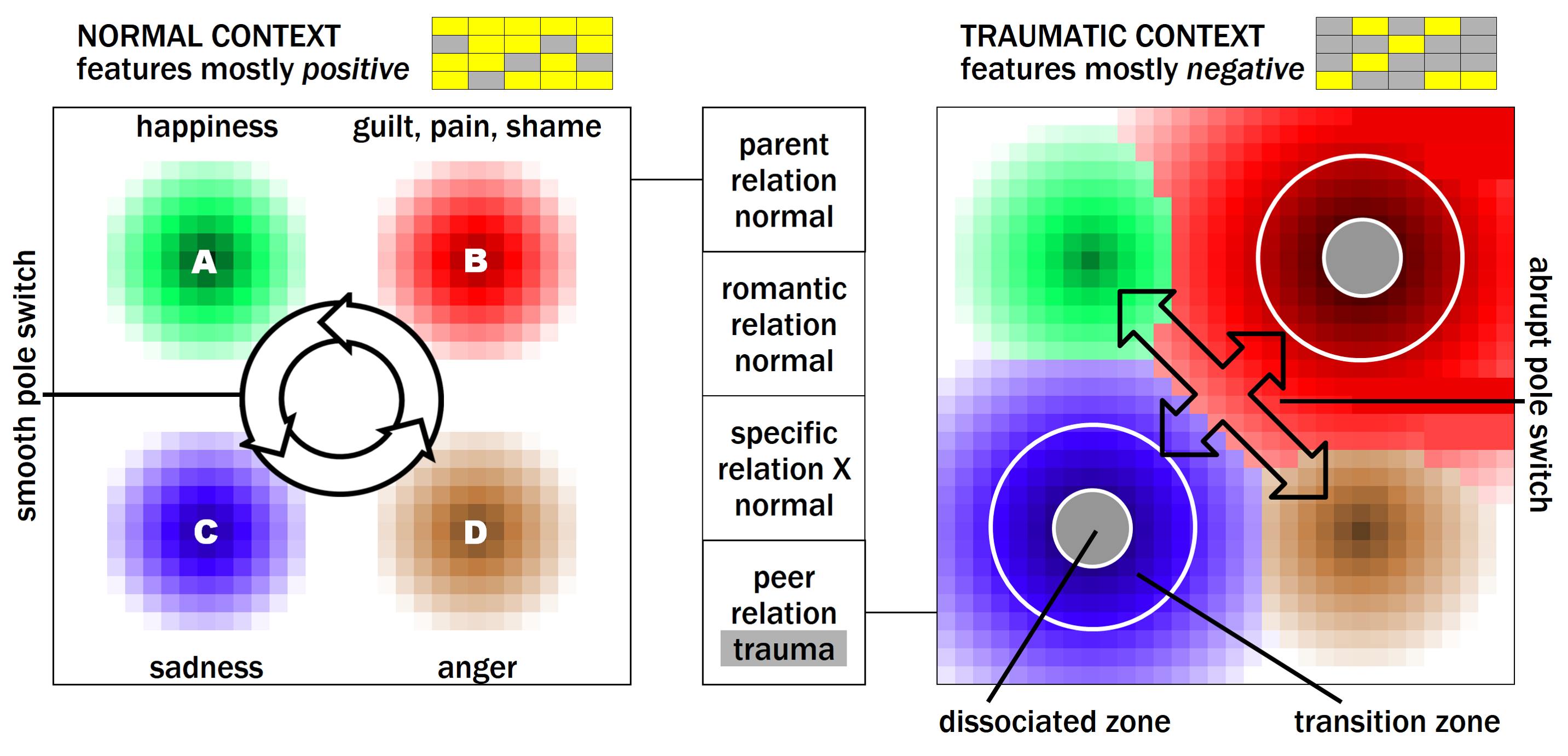}}}
\caption{Left: normal context. Each point in a context plane represents a situation (defined as a set of coactive features) and is associated to an emotion (whose intensity is represented by the colour shade). The mind can switch smoothly between emotional poles, no pole requires dissociation and the repertoire of emotions is fully accessible. Right: traumatic context. In a traumatic context, emotional poles B and/or C are characterised by too intense emotional levels and are inhibited. When the mind happens to be in one of these poles (forbidden zone), dissociation intervenes. To avoid dissociation, the mind oscillates between prototypical zones of permitted poles A or D, staying in each pole as long as the situation remains prototypical. When the mind nears a traumatic pole (transition zone), splitting symptoms appear and the mind switches to the prototypical zone of another permitted pole.}
\label{mapcon}
\end{center} \end{figure*}

A context can be conveniently represented on a plane, where each point corresponds to an individual situation (Fig.~\ref{mapcon}, left) and belongs to the ``zone of influence'' of an emotional pole, characterised by a specific emotion. Each pole originates from a point representing the most prototypical situation associated to the pole, and extends towards less prototypical situations. The epicentre of the anger pole, for instance, may correspond to a situation-point characterised by an object behaving very dishonestly, eliciting a very strong anger, while points further away may be characterised by a better behaviour of the object. We define a context ``normal'' if the highest emotional levels are not \emph{too} high. In this condition the mind can switch smoothly between all poles, experiencing different levels of the emotions associated to each pole.

In our definition, a \textbf{social trauma} occurs when the intensity of the elicited social emotion (Fig.~\ref{emotpoles}) is too high and exceeds the individual's tolerance threshold (this definition is somewhat broader than the one normally used and includes events that would be normally characterised as humiliating experiences). We assume that the emotional poles that can be involved in a trauma are those characterised by a negative self value: B (shame, pain, guilt) and C (disappointment, sadness). 

This in turn requires that the self-associated features have, on average, a negative value. If a person with bat ears, big nose and thin lips thinks that these features are very negative and s/he gets criticised or made fun of for them, a trauma may take place. Emotional poles with positive self value, such as love and happiness, but even anger and disgust, cannot become traumatic.  

The standard response to trauma is represented by the phenomenon of \textbf{dissociation}, defined as the distortion, limitation or loss of the normal associative links between perceptions, emotions, thoughts and behaviour. Dissociation can take the form of mental ``black-out'', depersonalisation (feeling of separation from one's body), derealisation (feeling of being detached from the world), selective amnesia and emotional detachment \citep{lanius2015, moskowitz2008, nijenhuis2011}. 

In our model, the space around a traumatic pole can be divided into three zones (Fig.~\ref{mapcon}, right): the ``forbidden zone'', an area around the centre in which strong, ``black-out''-like dissociation is used; the ``transition zone'', a safety belt around the forbidden zone; the ``free zone'', an area sufficiently far from the centre, where the mind can stay without any risk of experiencing unpleasant emotions.

\subsection{Context switch, splitting}  
\label{subsec:context_switch}

In case of trauma, the adoption of dissociation makes it possible for the mind the stay on a traumatic pole, excluding the awareness of intolerable thoughts and emotions from consciousness. However, the disconnection of aspects of reality may hide potential dangers and have a high cost. Therefore, the mind tries to avoid traumatic poles and heads towards non-traumatic ones, where the perception of reality is not restricted. 

Let us assume that, in a traumatic context, the mind is initially near the happiness pole (Fig.~\ref{mapcon}, right). The mind will stay around this pole as long as the conditions are prototypical, i.e. as long as the object relation is perfect, full of trust, mutual respect, etc. As the situation departs from the prototypical scenario of the gratitude pole and drifts into the transition zone of the disappointment pole or the pain pole, the mind switches abruptly to the prototypical zone of the anger pole. 

When the situation deviates from the prototypical scenario of the anger pole, the mind returns to the gratitude pole, or goes to another permitted pole, and the cycle repeats itself. This corresponds to the defence mechanism of \textbf{splitting}, defined as the inability to integrate positive and negative aspects of self and others, which results is a view of the world in ``black and white'' \citep{perry2013}. 

The transition zone is a safety belt built around the forbidden zone, characterised by high emotional levels. This causes an increase of the emotional salience of the features involved in the trauma (link sx in Fig.~\ref{modules}), which represent the person's defects criticised. As a result, these features become the focus of attention and appear magnified and distorted: if the feature criticised is ``bat ears'', when looking at the mirror the person will see his/her ears magnified and more protruded, like in a caricature. The magnification of defects serves the purpose of warning the executive function that the current situation is close to a traumatic point, and gives an indication of the features that need to be closely monitored. 

This phenomenon is reminiscent of symptoms relevant to a wide range of mental disorders, starting from the hallucinations, the delusions and, more in general, the ``delusional atmosphere'' that characterise schizophrenia \citep{moskowitz2008}. Transitory perceptual distortions or ``pseudo-hallucinations'' are not uncommon also in personality disorders \citep{gras2014}. Perceptual distortions are present in body dysmorphic disorder, which can either appear stand-alone \citep{phillips2004} or be responsible for the dysmorphic body image associated to eating disorders (among many other conditions) \citep{ruffolo2006}. 

\subsection{Selective dissociation}  

\begin{figure*}[t] \begin{center} \hspace*{-0.0cm}
\includegraphics[width=17.00cm]{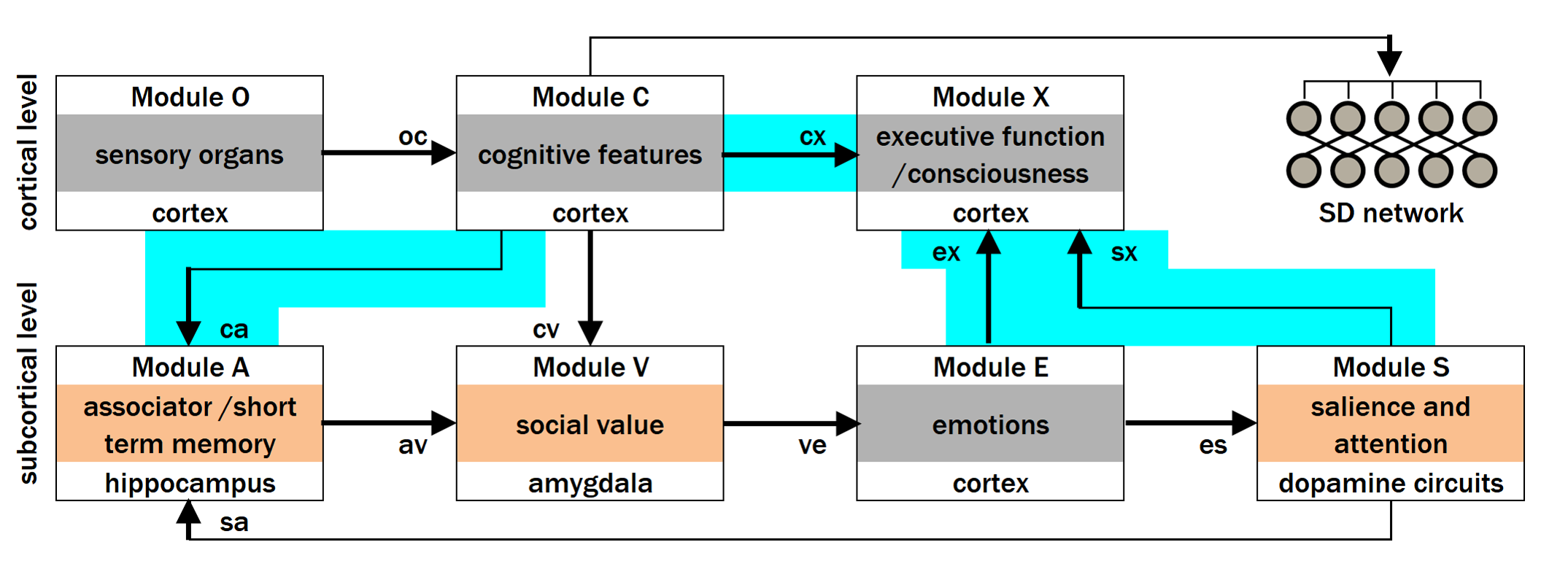}
\caption{Selective dissociation network. This figure shows the modules of Fig.~\ref{modules} with the addition of the SD (selective dissociation) network. The SD network receives in input cognitive features and send modulatory signals to cognitive and emotional links (shown in light blue). The effect of modulation can be perceived as a restriction of consciousness and, if the input to the associator (hippocampus) is affected, can block the storage of some features to short- and long-term memory.}
\label{sdnet}
\end{center} \end{figure*}

\begin{figure*}[t] \begin{center} \hspace*{-0.00cm}
\includegraphics[width=13.00cm]{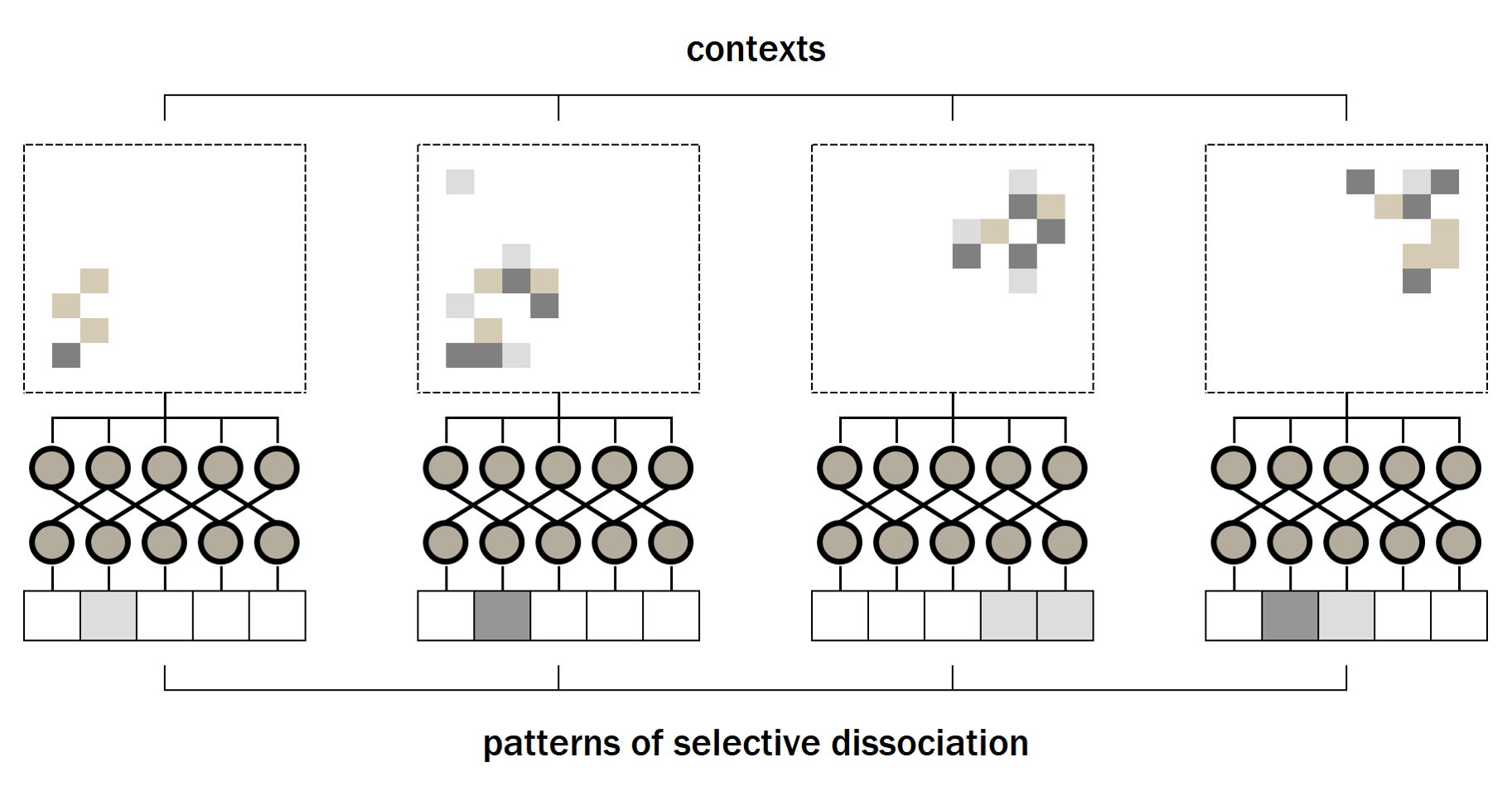}
\caption{Attribution of patterns of selective dissociation to contexts. The SD network classifies the input composed of cognitive features into a number of classes. For each class, it produces a specific pattern of selective dissociation, i.e. a combination of attenuation levels for cognitive and emotional features. Depending on the number of available neurons, the SD network can learn to map a different number of classes.}
\label{dissocpat}
\end{center} \end{figure*}

We can hypothesise that, upon the first occurrence of a trauma, dissociation is total and involves all mental functions: this may correspond to the ``freezing'' behaviour of ``disorganised attached'' children \citep{main1986}. Total dissociation provides a shield from pain, but may expose the individual to serious consequences in a potentially dangerous environment. For this reason, it is not unrealistic to assume that, in case of repeated traumas, the mind will try to replace total dissociation through less pervasive forms of dissociation, able to preserve a reasonable level of functioning.

This is achieved through \textbf{selective dissociation}, a limited form of dissociation which, near a traumatic pole, excludes from consciousness \textit{some} emotional or cognitive channels, leaving total dissociation active only in the forbidden zone. This is a common phenomenon. Medical students may be shocked when they see an operation for the first time, but they rapidly get used to it. The same happens to the medical staff of intensive care units, where death is a daily occurrence, or to the employees of a slaughterhouse. This happens to all of us, when we see poor people living in the street and pretend it is normal.

We can consider the process that leads to selective dissociation as a form of learning, in which the mind selects the smallest subset of reality that needs to be excluded from consciousness to avoid unpleasant emotions without losing touch with the ``here and now''. In its simplest form, it is obtained through a reduction of the ``volume'' of painful emotions, but it can also involve cognitive distortions.

This definition implies that normal and pathological dissociation are manifestations of the same tool adopted by the mind to filter and to adapt the information flow to consciousness and to memory: the defence mechanism used by the doctor to detach the emotions associated to illness and death is the same that causes inhibition of mentalisation, by which the victim of abuse becomes unaware of the malevolent intentions in the abuser's mind \citep{fonagy2002}, or multiple personalities. Both normal and pathological dissociation are adaptive phenomena, installed in the mind by Darwinian evolution to provide the best compromise in difficult situations. 


It would be interesting to explore the interactions between selective dissociation and the phenomenon of habituation, defined as a reduced response to a repeated stimulus \citep{rankin2009}. We might hypothesise that, while habituation mostly refers to neutral stimuli (such as, e.g., a harmless loud noise), selective dissociation intervenes to reduce the intensity of negative emotions. A thorough analysis of the topic is beyond the scope of this work, but definitely in the agenda of future developments.

The learning of selective dissociation is obtained through a modulation of the links between mind modules, which could be attenuated in certain situations. We hypothesise that this is achieved through a dedicated \textbf{selective dissociation (SD) network} that receives in input the set of cognitive features and send modulatory signals to emotional and cognitive links (Fig.~\ref{sdnet}). In other words, for each situation, the network decides the degree of attenuation of each channel. As a result, different classes of situations are selectively mapped to different attenuation patterns (Fig.~\ref{dissocpat}). 
 
The effect of modulation can be perceived as a restriction of consciousness and, if input to the associator (hippocampus) is affected, can block the storage of some features to short- and long-term memory. The effect of selective dissociation is a ``restructuring'' of the space around traumatic poles, with an overall reduction of emotional levels and a marked reduction of transition zone and forbidden zone. 

With selective dissociation, the mind can once again switch through all emotional poles. The difference between a normal condition and a pathological one lies in the extent of transition zone and forbidden zone (in which total dissociation must be used). If these are large, the pole switching can still happen but is not fluid: in this case we speak again of (mitigated) splitting. 

The restriction of consciousness imposed by selective dissociation is a well-known phenomenon. On the other hand, there is ample evidence for the existence of subconscious, dissociated memory records, which are denied access to consciousness, but which can re-emerge after a period of time, when the conditions that justified their suppression have been removed. Painful memories, such as incest or domestic violence, can be reactivated after many years, when the person feels safe (e.g. in a therapeutic setting). 

For these reason, we hypothesise that the records stored in memory are always complete (including the dissociated components). Once the record is recalled from memory, such components are again dissociated and excluded from consciousness, until the SD network configuration change. The exception to this rule is when selective dissociation affects also incoming connections to the associator (hippocampus), in which case the information is not recorded and cannot be retrieved. 

\subsection{Malfunctioning of dissociation} 

The SD network can exhibit a degraded level of functioning for a number of reasons. One possible reason is that the SD network is endowed with an insufficient number of neurons and/or connections. Another possibility is that, albeit the connectivity of the SD network is normal, its functioning is impaired by other factors, that perturb the physiology of neurons and/or that of auxiliary cells. As a result, the classification ability of the SD network is poor: instead of providing a customised dissociative pattern for each trauma, the network is only able to generate a limited set of distinct patterns, leading to an insufficient attenuation of trauma-induced emotions and salience levels.

The root cause of the impaired functioning of the SD network can in turn be genetic (gene variants prevent neurons and/or connections to be generated or correctly assembled during embryonic development) or environmental (events occurring during brain development influence the rate of generation of neurons and/or connections). Possible events include difficulties during birth, abuse of drugs, exposure to chemical substances during gestation, etc. Another possible cause of alteration of neurons' operations is represented by inflammation and/or by an anomalous functioning of the immune system, which seems to be implicated in a broad range of mental conditions.

The central role that in our model the SD network is hypothesised to play in all physiological and pathological conditions, explains why the factors that contribute to its malfunction are shared among a broad set of mental disorders, and why they can lead to one disease (e.g.: schizophrenia) in a family member of one generation, and to another disease (e.g.: bipolar disorder) in a family member of the next generation. The diversity of such factors, on the other hand, can be explained by the Anna Karenina principle, which states that ``all happy families are alike; each unhappy family is unhappy in its own way''.


\newpage
\section{Trauma and complex trauma}  
\label{sec:trauma}

\subsection{PTSD}  

Post-Traumatic Stress Disorder (PTSD) is a condition caused by the exposure to one or more stressful events of extraordinary magnitude or catastrophic nature. Typical PTSD symptoms usually appear within a short period of time after the traumatic event; they include: hyperarousal, re-experiencing of traumatic memories (the so-called flashbacks), emotional numbing, helplessness and a shaking of one's self and perception of the world. The traumatic experience can have different outcomes in different persons: some people fully recover within a short period of time, while others go on to develop PTSD. 

Based on our definition, a psychological trauma is caused by a ``bad'' emotion which overwhelms the coping ability of the mind. The list of bad emotions includes the fear of dying and the pain associated to sexual rejection, but in this work the attention is focussed on social emotions: as a result, we restrict the set of emotions which can potentially cause a trauma to guilt, shame, social pain (Fig.~\ref{emotpoles}). Strong bad emotions are in turn caused by negative self-associated cognitive features: in the case of simple PTSD, the traumatic landscape is characterised by few very negative features in few contexts (most contexts are unaffected). Dissociation works normally (no genetic or neurodevelopmental issues) and selective dissociation is deployed in traumatic contexts, but is unable to fully patch up the trauma(s). 

A typical anatomical characteristic of PTSD is represented by structural alterations to the hippocampus \citep{bremner2006}. Since the hippocampus is the device that provides the initial association between co-occurring features, it is natural to think that this structure is also involved in the absence of association. Dissociation translates to a negative modulation of the information fed to the hippocampus and presented to consciousness, through links ca and cx in Fig.~\ref{modules} which, based on our model, are the carriers of cognitive features. This leads to a shrinkage of their target areas, which determines in turn a reduction in hippocampal volume. 

Flashbacks are elicited by situations similar to that of the trauma. Selective dissociation diverts aspects of reality from consciousness and memory, which contributes to render the contexts in which flashbacks appear less specific and hence more pervasive. If the awareness of war is dissociated, the sound of bombs is indistinguishable from the sound of firecrackers or fireworks; if the malevolent intention in the abuser's mind is dissociated, the abuser will look like any other man or woman. As a result, a larger set of situations will be feared and avoided. 

The variable severity of the traumatic disorder is a phenomenon can be accounted for in our model. When the trauma occurs, e.g. when a woman is raped, a new negative feature is linked to the person's self, i.e. ``being a raped woman'', causing a sudden drop in the person's self esteem. In the weeks afterwards, this feature enters in contact with other self-associated features. If these are mostly positive (e.g. ``being a good mother'', ``being an esteemed university professor''), the positive value flows from these old features to the newly acquired feature, which becomes less negative, until the trauma is solved. If on the other hand the old features are mostly negative (as a result of the person's past experiences), this healing process cannot occur, negative features reinforce each other and the trauma persists. 


\subsection{Complex PTSD}  

Complex PTSD is a mental condition caused by severe and persistent and /or repeated trauma (maltreatment, sexual abuse, physical or emotional neglect in childhood, disfunctional relationships in adulthood, etc.). In contrast to classical PTSD, in complex PTSD traumas are repeated over a long period of time. The outcome is characterised by a broad spectrum of cognitive, affective and psychosocial impairments, which usually have a long lasting impact on the individual's life.

The main characteristic of complex PTSD is the duration of exposure to traumatic events, which often occur during the entire childhood and beyond. As a result, according to our model, there will be many negative features in the patient's mind, for two reasons. On one hand, the negative ideas directly originating from the trauma (e.g., ``being sexually abused by your father'', ``being neglected by your mother'', ``being bullied by your schoolmates'') get reinforced and solidified through repeated exposure. This leads to a crystallisation of such negative ideas, that become part of the person's self.

On the other hand, after the original negative ideas have become steadily associated to the self, they transfer negativity to other neutral self-associated ideas such as, e.g., ``being a baseball player'', when negative and neutral ideas happen to co-occur in the individual's mind (e.g., the person remembers being bullied at school while playing baseball). As postulated, value propagates by association and, as time goes by, this mechanism generates a large set of negative self-associated ideas, that keep re-occurring, thereby reinforcing each other. 

The SD network, which is hypothesised to function normally, intervenes and deploys customised dissociation patterns in each context, which contributes to alleviate the emotional stress, making the situation more tolerable. However, given the breadth of contamination of negative ideas in the individual's mind, selective dissociation is unable to fully patch up the traumas. For all these reasons, complex PTSD is a condition much more stable and difficult to solve, compared to standard PTSD.

\subsection{Dissociative Identity Disorder}  

Dissociative Identity Disorder (DID) is characterised by the presence of different personality states (dissociative identities) which take in turn control of the thinking, feeling and acting of a person. These different personalities have their own characteristics, behaviours, abilities, patterns of perception and thinking. A key DID symptom is the evidence of memory gaps relevant to events or personal information that cannot be explained by ordinary forgetfulness.

Like complex PTSD, also DID is the usually the outcome of a long exposure to traumatic events. As a result, many negative features are present in the person's mind, and many contexts are involved. As in the case of complex PTSD, the broad presence of negative ideas is determined by the propagation of negativity by association: once the set of negative ideas reaches a critical mass, the contamination becomes difficult to stop. The resulting situation corresponds to a stable equilibrium point, whose attraction is difficult to resist.

The SD network reacts by deploying customised dissociation patterns, in a large number of contexts. The main characteristic of DID is the peculiar relation between the different customised dissociation patterns, used in as many traumatic contexts. While in complex PTSD the various dissociative patterns have much in common (the subset of non-dissociated cognitive features is large), in DID this set is much smaller. As a result, the continuity of the self is lost and the different dissociative patterns behave as independent personalities.

\clearpage

\newpage
\section{Psychosis} 

\subsection{Schizophrenia}  

Schizophrenia is often described in terms of ``positive'' and ``negative'' symptoms. The positive symptoms are those that do not occur normally in healthy people: they include delusions, and tactile, auditory, visual, olfactory and gustatory hallucinations, generally considered as manifestations of psychosis. The negative symptoms correspond to deficits of normal emotional responses or other thought processes: they include flat affect, lack of emotions, poverty of speech (alogia), inability to feel pleasure (anhedonia), lack of desire to form relationships (asociality) and lack of motivation (abulia).

Schizophrenia is a complex condition in which both genetic and environmental factors play a role. The heritability of the disease is very high, and GWAS studies identified a large number of gene variants correlated with the disease, each present in a small fraction of cases. A recent genetic study \citep{sekar2016} found a correlation between schizophrenia and a gene coding for a receptor that targets synapses for destruction by the immune system: according to the authors, psychosis would be caused by excessive synaptic pruning. As already pointed out, the immune system seems to play a role in a wide range of mental conditions, including schizophrenia.

A history of childhood traumas is reported to be a risk factor for schizophrenia. The nature of positive symptoms (e.g. the content of auditory hallucinations) is consistent with low self-esteem and negative affect, which are the common outcome of adverse childhood experiences \citep{schaefer2011}. A theory that puts in evidence the role of trauma at various levels is the ``traumagenic neurodevelopmental model of schizophrenia'' \citep{read2014}. The interplay between genetics and trauma remains elusive.

According to the so-called ``dopamine hypothesis'' \citep{howes2009dophyp}, psychosis is caused by ``aberrant salience'', determined by an excess of neurotransmitter dopamine in the brain. This hypothesis is supported by the fact that antipsychotic drugs act by blocking the D2 dopamine receptor.  However, is was not possible to identify schizophrenia-associated genes involved in dopamine transmission \citep{alexis2016}. 

Our model has already be used to interpret schizophrenia \citep{fontana2017schizo}. As a result of trauma, many contexts, such as ``parent relation'' and ``peer relation'', are traumatic. To avoid the forbidden zones of traumatic poles (Fig.~\ref{mapcon}, right), the mind is forced to oscillate between permitted poles (splitting). In the transition zone, as described in section \ref{subsec:context_switch}, the features involved in the trauma are overloaded with salience and appear magnified and distorted, like in a caricature: we call such distortions ``pseudo hallucinations''. 

When the mind gets too close to the epicentre of a trauma, selective dissociation should be deployed, to remove disturbing thoughts and turn down the ``volume'' of bad emotions. Our hypothesis is that the learning process that leads to selective dissociation is defective in schizophrenia: dissociation is either too weak or too strong. When it is too weak, painful emotions are not attenuated, and the perceptual distortions caused by excessive salience proceeds unabated: this is, in our opinion, the nature of hallucinations. 

Dissociation eventually intervenes, too strongly, producing the full spectrum of negative symptoms: emotional detachment, anhedonia, lack of motivation. Likewise, dissociation is responsible also for cognitive symptoms: a number of cognitive features are prevented from reaching the consciousness and accessing memory, and the mind has a limited repertoire of ideas to use. This causes poverty of speech and thought disorganisation. The degree of dissociation correlates with the severity of symptoms.

\begin{figure*}[t] \begin{center} \hspace*{-0.00cm}
\includegraphics[width=16.00cm]{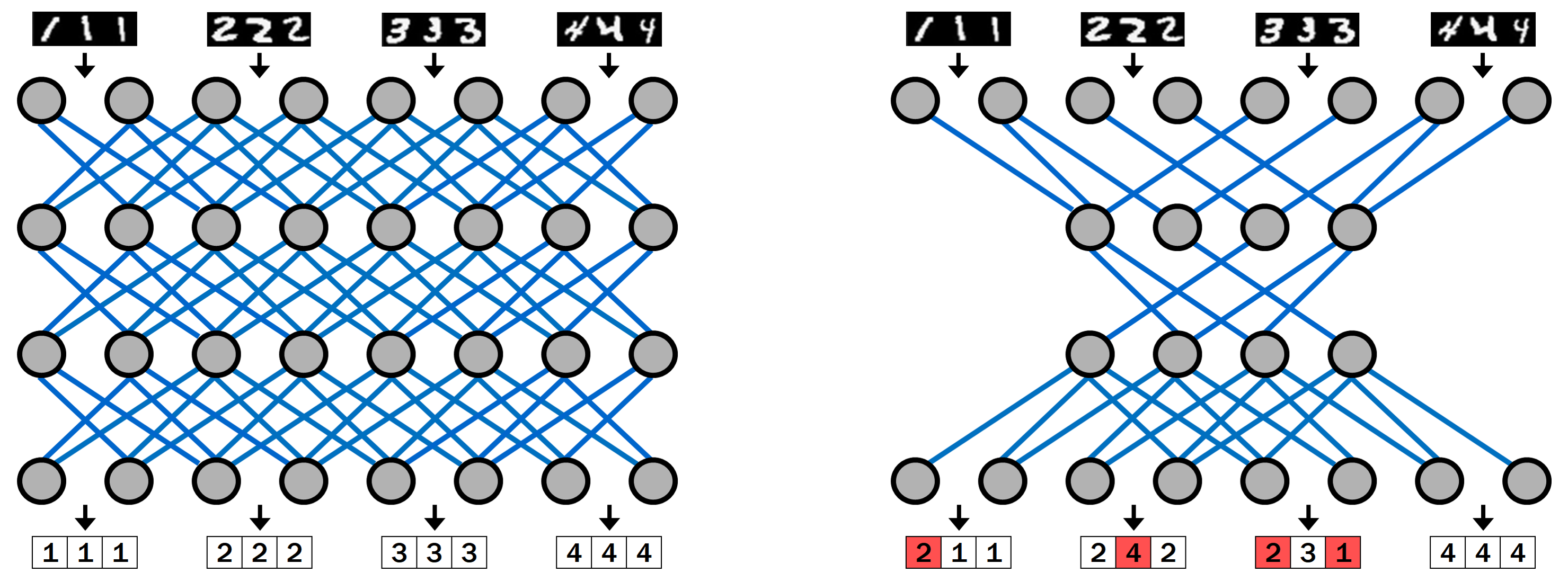}
\caption{Left: neural network used for visual recognition. The network receives in input the image of a handwritten digit and produces in output a number corresponding to the digit. Right: a neural network with fewer neurons and/or connections has a more limited classification power and is unable to correctly classify all inputs (errors are marked in red).}
\label{neurals}
\end{center} \end{figure*}

To understand how the SD network can be defective, we can look at artificial neural networks, which can be used for classification tasks, such as the classification of handwritten digits (Fig.~\ref{neurals}). In this case, the network receives in input the image of the character and produces in output an integer number in the [0,9] range representing the class. In order to effectively solve a given classification task, a neural network must have a sufficient number of neurons and/or connections between neurons, otherwise the classification will be inaccurate, e.g., a ``2'' will be misclassified as a ``4'' (Fig.~\ref{neurals}). The number of neurons required depends on the complexity of the task (i.e., the number of classes to be discriminated).

Our conjecture is that, in schizophrenia, the SD network has an insufficient number of neurons and/or connections. As a result, its classification ability is poor: instead of providing a customised dissociative pattern for each context, the SD network is only able to generate a few distinct patterns. For some situations the dissociative pattern will be too weak, causing positive symptoms, for other situations it will be too strong, causing cognitive and negative symptoms.  

The dopamine hypothesis maintains that hallucinations and delusions are caused by an excess of salience, in turn driven by an excess of dopamine in the limbic system. However, based on our model, the excess of dopamine is not due to a malfunction of the salience module, but it is caused by too intense emotional levels, left undissociated by a defective SD network. The block of dopamine receptors, obtained with antipsychotic drugs, translates to shutting down the salience mechanism in all situations and leads to an abnormal functioning of the brain, which may exacerbate the negative symptoms.     

The SD network can be thought to perform an inhibition and control function, a role that pertains to the prefrontal cortex. As a result, the hypothesised reduced number of neurons and/or connections in the SD network could be correlated to the structural changes frequently observed in the prefrontal cortex of schizophrenia patients \citep{hugdahl2015}. Also the hippocampus of schizophrenic subjects is significantly smaller and seems to be less prone to the phenomenon of habituation \citep{williams2013}, which could be needed to develop selective dissociation.

Besides genetics, several environmental, non-trauma-linked factors, such as difficulties during birth, use of drugs or inflammation, are correlated with schizophrenia. Our hypothesis is that all these factors act by reducing the number of neurons and/or connections in the SD network, building a predisposition that requires the action of traumas to be realised. The accumulation rate of traumatic events is expected to be constant in the course of life and to accelerate during adolescence, which is acknowledged to be a stressful period: this would explain why schizophrenia has its typical onset at the end of adolescence. 

\subsection{Bipolar disorder}  

\begin{figure*}[t] \begin{center} \hspace*{-0.00cm}
\includegraphics[width=17.00cm]{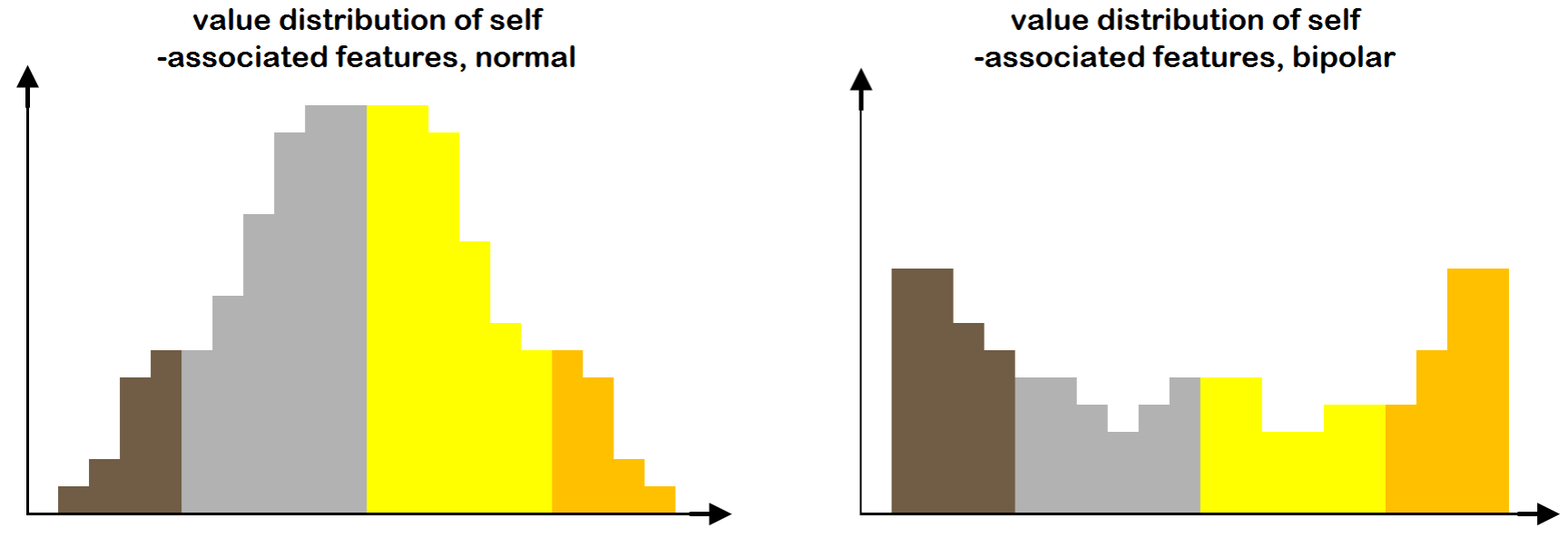}
\caption{Statistical distribution of feature values. On the left: histogram of feature values for a normal condition. Features have a Gaussian distribution, where most features have a value close to the mean and few extreme points are present. On the right: histogram of feature values in case of bipolar disorder. The standard deviation is much higher and many features with extreme values are present.}
\label{histofeat}
\end{center} \end{figure*}

Bipolar disorder (BPD) is characterised by an alternation between two opposite emotional states, one dominated by excitement (the so-called mania) and one by depression. The ``core'' mood disorder is often accompanied by alterations in thought processes, motor skills and behavior, as well as by neurovegetative manifestations (anomalies of energy levels, appetite, sexual drive, sleep-wake cycle). A first distinction is made between unipolar and bipolar disorder. A further distinction can be made between bipolar I disorder (characterised by manic episodes) and bipolar II disorder (characterised by hypomania and recurrent depressive episodes).

Twin studies have established that bipolar disorder is, along with schizophrenia, one the most heritable medical disorders, with genetic influences explaining 60-85\% of risk \citep{Smoller2003, Barnett2009}. GWAS studies have shown that the effect size of BPD-associated gene variants are modest, suggesting that the genetic basis of BPD is likely to be determined by polygenic effects of many genes, each with a small contribution to predisposition. 

Personality traits and environmental influences play a crucial role in the aetiology of BPD. Stressful life events such as traumatic experiences  or psychosocial stress can trigger disease phases \citep{Aas2016}. Triggering factors include a lowering of self-esteem, an irregular lifestyle, abuse of alcohol and/or other substances. Up to 75\% of patients report to have experienced intense stress immediately before the first BPD episode.

Dopamine is hypothesised to play a role also in bipolar disorder \citep{ashok2017}. Evidence from pharmacological and imaging studies support the hypothesis that a state of hyperdopaminergia (elevated D2/3 receptor availability) is associated to manic episodes. Strangely, both dopamine agonists and antidopaminergics seem to improve BPD symptoms. The first choice drugs in the treatment of BPD are lithium salts (usually lithium carbonate). 

Based on our model, we can hypothesise that in future BPD subjects (as well as in normal subjects), negative ideas keep accumulating during childhood and adolescence. Dissociation follows-up the situation and builds tailored dissociative patterns for each context. The outcome of this process is that in such contexts (some) negative features are inhibited from reaching consciosness. With this premise, we can imagine two mechanisms for the interpretation of BPD.

The first mechanism is based on the assumption that dissociation works normally and traces back the origin of BPD to the structure of the set of features. Mood has previously been defined as the long term self value, determined by the features associated to the self in a stable way (e.g. profession, curriculum vitae, wealth, etc.), as opposed to features that change on a more rapid timescale (e.g. whether last night I won the tennis match with my friend or not). Such features become part of the subject's identity.

If all features are characterised by a not too negative social value, their activation or deactivation does not cause large mood oscillation. If, on the other hand, there are very negative features (e.g. I have a big nose and I think it is a really bad thing), the mood will be very low when the feature is present, and very high when it is dissociated. The situation is analogous to that of a balloon to which a heavy weight is attached: once the weight is removed, the balloon makes a leap upwards. This could explain the large swings typical of bipolar disorder. In other words, the bipolar mood would correspond to a bipolar distribution of feature values (Fig.~\ref{histofeat}). 

The second mechanism, instead, is based on a malfunctioning SD network. Let us assume that the mind is intially in a bad context, that we might call ``depression context'', characterised by many negative features, removed from consciousness by selective dissociation. Let us now imagine that a switch takes place to a new context, that we will call the ``happiness context'', dominated by positive features and pleasant emotions. In the new context, the dissociation of negative features is not needed and should be disabled.

Let us now hypothesise that the dissociative pattern does not change, but remains that of the depression context. This leads to a condition in which only positive features are present and creates the premises for a manic episode. In other words, mania is a state in which negative features are dissociated and only positive features are allowed to access consciousness. Once the context switches again back to the one dominated by negative features, an analogous mismatch of dissociation prevents the attenuation of negative emotions and leads to depression.

The inability of the SD network to adapt the dissociation level to the context is determined, as in the case of schizophrenia, by a reduced number of neurons and/or connections, caused by the same genetic, neurodevelopmental and environmental factors. This explains 1) why the genetic contribution is spread on a wide range of genes and gene variants and 2) why such genes are shared between the two conditions and can lead to one disease in a family member of one generation, and cause the other in a family member of the next generation.

\clearpage



\newpage
\section{Therapeutic proposal}

\subsection{Theoretical foundations}

As pointed out in section \ref{sec:modelmind}, the intensity of emotional pain is inversely proportional to the self value and directly proportional to the object value, which in turn depend on all associated features. If the self value is negative, it is because most self-associated features are negative; if the object value is positive, it is because most object-associated features are positive. It is impossible to have a negative self linked to mostly positive features, or a positive object linked to mostly negative features. 

The sets of features associated to self and object represent the foundation of the traumatic structure. If the values of such features are modified, the values of self and object are expected to change accordingly and, as the value gap narrows, the level of emotional pain is expected to diminish, until the resolution of trauma. Therefore, our therapeutic strategy is to target self- and object-associated features and change their values in the patient's mind. As we have seen, the good news is that value is attributed to most features in a completely arbitrary way. The value background of a person is initially set by the senior family members (usually the parents), but in principle nothing prevents to bring changes to it.   

Following these principles, our therapeutic method aims to bring changes to the value system through \textbf{counterexamples}. Assuming, for instance, that the feature ``taking risks'' has a negative value, we can provide counterexamples in which a risk-taking behaviour gave good results. We might mention Julius Caesar, who chose to cross the Rubicon; Butch Cassidy and Sundance Kid, who jumped into a waterfall and saved their lives (at least in the movie!). Given the negative feature ``failing'', we might say that Henry Ford was bankrupt a number of times before succeeding, and so on. Linking positive examples to negative features raises the features' value. Likewise, the association of negative examples to positive features can be used to lower the features' value. We call this therapeutic scheme \textbf{Value-Oriented Counterexample-Supported Massively Associative Training (VOCSMAT)}.

From the neurobiological point of view, our hypothesis is that the functioning of the emotional module (Fig.~\ref{modules}) is genetically determined and not plastic. In other words, given the features' values, the type and intensity of the emotion generated are completely determined. The plasticity lies instead in the associator and in the social value module: the features' values can be modified by recording new associations in the hippocampus, which in turn affect the coding of values in the amygdala (Fig.~\ref{hippamigd}).

Since ``negativity'' propagates by association (and is very contagious!), there might be many (several hundreds) features linked to traumas in the patient's mind: such features need to be \textit{all} addressed. For each feature, usually 3-4 counterexamples may be needed, which brings the total number of counterexamples around the figure of 1000. Finally, the exposure to data has to be repeated multiple times, over a period of several months. 

The set of 1000+ examples to be used represents the psychological analogous of ``big data'', although in the case of neural networks the number of examples needed can be higher by several orders of magnitude. The reader may wonder why the technique of learning through examples, that works for neural networks, would work also for humans. The answer is that this machine learning technique is a mechanised form of a kind of learning that is performed also by biological brains: it is thus straightforward to conclude that it can (and must) work for humans too.    


One of the expected effects of the therapy is a ``restructuring'' of dissociation,  with a marked reduction of forbidden and transition zone. The effect of dissociation can be likened to the action of weirs that separate the healthy parts of the mind in water-tight compartments. When a ``bad'' compartment is excluded, the rest of the mind /memory is made immune to the flooding of negative features and bad emotions. Once the feature space is restructured through therapy, bad emotions become less bad and more tolerable and the weir reopens: when this happens, we can expect a sudden decrease of self value and a flood of bad emotions. However, the effect is temporary and marks the start of the integration process.

\subsection{Implementation considerations}
  
A potential problem we can expect to encounter, due to the large number of features involved, can be clarified through an example. Let us suppose to have 100 black socks that we want to turn into white socks, by treating a given number of socks at a time with a whitening solution. Let us also suppose that, at regular intervals, a random subset of socks is put into a washing machine. If the white socks in the washing machine are much fewer than the black socks (as we can expect at the beginning of the treatment), they will absorb the colour released from the black socks and turn black again. Only when the white socks become the majority, is the ``washing machine effect'' expected to be beneficial, but this may take a long time. 
 
In this metaphor, black socks correspond to negative features, white socks correspond to positive features and the whitening solution corresponds to the VOCSMAT treatment described above. The washing machine corresponds to the associative mechanism of the mind, that continues to function during the therapy. If the therapy could be administered ``offline'', there would be no washing machine, but this is not possible. The therapy has to be carried out ``online'', with the mind fully connected to the external world, which might remind the patient how unworthy he/she is.

The reader might argue that the successful application of this technique require that the patient be aware of his/her emotions, of the features involved in a trauma and their values. On one hand, due to dissociation, such awareness is not guaranteed to be present. On the other hand, this requirement is not needed, as the counterexamples will be administered by the therapist. It is to be expected, especially in the first phases of the therapy, that some counterexamples will not be consciously perceived by the patient's mind, due to dissociation and other defence mechanisms. As the therapy progresses, the patient is expected to become increasingly receptive.  

The VOCSMAT therapy has some points of contacts with CBT: the key difference between the two can be illustrated with an example. Let us assume that the patient is a teenager boy, with hostile parents and no friends, convinced that these features are negative and that he is a loser. CBT could try to convince the patient, with logical arguments, that he possesses other qualities that the others can appreciate, that there are people who love him, etc. In reality, the patient may be right in thinking that most people think he is a loser: in this case CBT tries to prove something untrue and is doomed to failure.

The VOCSMAT therapy does not try to enhance and put in evidence the positive features of the patient (it is not needed). It accepts reality as it is and focusses on the negative features, trying to raise their value. If the patient says: ``I'm a loser, my mother is angry at me, I don't have friends at school, everybody makes fun of me and neighbours spy on me'', the VOCSMAT therapist may say: ``Ok. So what? Even Albert Einstein was a loser at school, even Napoleon had no friends, even George Washington had a mother who ... etc. You have mental problems? You cut yourself and have other weird behaviours? You are not alone, many artists and scientists had mental problems: Charles Bukowski, John Nash, Vincent Van Gogh ...'' and so on. In this way the patient begins to value himself for what he really is in the present, and not for what he might become in the future.

The reference to the present situation of the patient might remind some aspects of Gestalt therapy \citep{Brownell2010}, a method focussed upon the individual's experience in the present moment, aimed to raise awareness (also called ``mindfulness'' in other clinical domains), based on the assumption that perceptions, feelings and acts are conducive to interpretation and conceptualisation. This framework has in fact little to do with the VOCSMAT therapy, that is not concerned with perceptions, feelings or situational awareness, but is focussed on changing the value of the individual's negative features in the present as well as in a hypothetical future. 

Features can be divided into three main categories: 1) features that do not depend on the self and that can occur at any time (present and future), e.g.: ``being criticised''; 2) features that depend on the current self, e.g. ``not taking initiatives for fear of criticism'' (this category also includes pathological symptoms, e.g. risky behaviour, substance abuse, etc.); 3) features associated to an ideal self (in the future), for example: ``undertaking risky initiatives, daring''. As seen from the example, it is possible to have a feature in the second category and its opposite in the third: the value of both features must be raised (the rules of arithmetic do not apply in this case!). The idea to address features linked to the current self is indeed compatible with Gestalt's ``paradoxical theory of change'', according to which ``change comes about as a result of full acceptance of what is, rather than a striving to be different'' \citep{Beisser1970}.


\subsection{Examples of features linked to traumas and counterexamples}

This section reports lists of features related to ten (common) traumas (for the last four counterexamples are also provided): 

\begin{table*}[h!]
\vskip 0.00cm
\centering
\begin{tabular}{l l}
1) ``Abandonment'' (emotion: pain)            &  2) ``Being bad'' (emotion: guilt)              \\
3) ``No consolation'' (emotion: sadness)      &  4) ``Being controlled'' (emotion: pain, shame) \\
5) ``No recognition'' (emotion: pain, shame)  &  6) ``Being rejected'' (emotion: pain, shame)   \\
7) ``Making mistakes'' (emotion: pain, shame) &  8) ``Being weak'' (emotion: pain, shame)       \\
9) ``Being different'' (emotion: pain, shame) & 10) ``Being a loser'' (emotion: pain, shame)    \\
\end{tabular}
\vskip 0.00cm
\label{words}
\end{table*}

Interestingly, many lists are present on the internet, that link negative features to famous (hence positive) persons, e.g.: ``Celebrities who were bullied before becoming famous'', ``Gorgeous stars who were dumped in the worst ways possible'', ``Famous people with disabilities'', etc. Internet users seem to be fond of such lists because, we think, these associations raise the value of self-associated negative features in the users' minds. As a consequence, the self value improves and these persons feel better.

\clearpage \pagebreak[4]

\begin{figure}[p] \begin{center} \hspace*{-0.25cm}
{\fboxrule=0.0mm\fboxsep=0mm\fbox{\includegraphics[width=17.50cm]{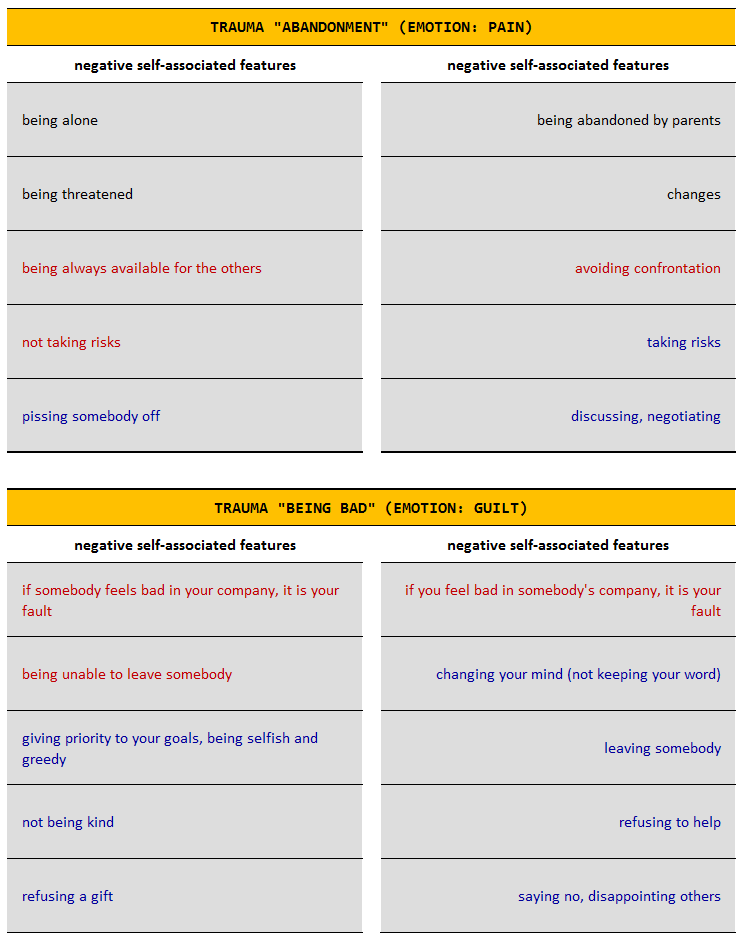}}}
\caption{Some features related to traumas ``ABANDONMENT'' and ``BEING BAD''. Features coloured in black belong to the first category (independent from self), features coloured in red to the second (linked to current self), features coloured in blue to the third (linked to future self).}
\label{abandon}
\end{center} \end{figure}

\begin{figure}[p] \begin{center} \hspace*{-0.25cm}
{\fboxrule=0.0mm\fboxsep=0mm\fbox{\includegraphics[width=17.50cm]{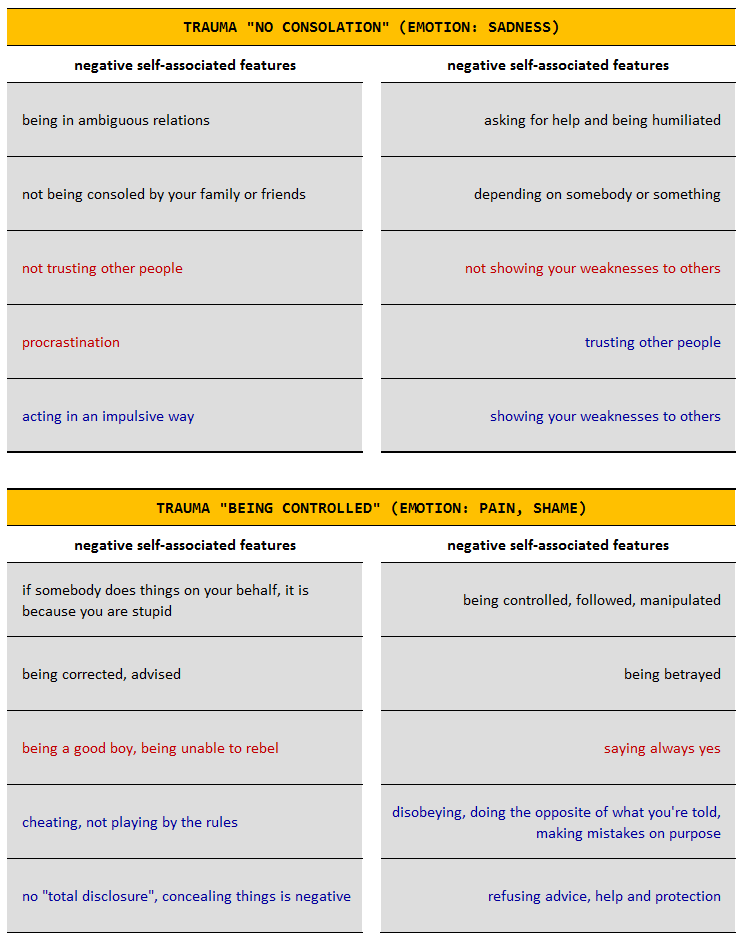}}}
\caption{Some features related to traumas ``NO CONSOLATION'' and ``BEING CONTROLLED''. Features coloured in black belong to the first category (independent from self), features coloured in red to the second (linked to current self), features coloured in blue to the third (linked to future self).}
\label{nocons}
\end{center} \end{figure}
 
\begin{figure}[p] \begin{center} \hspace*{-0.25cm}
{\fboxrule=0.0mm\fboxsep=0mm\fbox{\includegraphics[width=17.50cm]{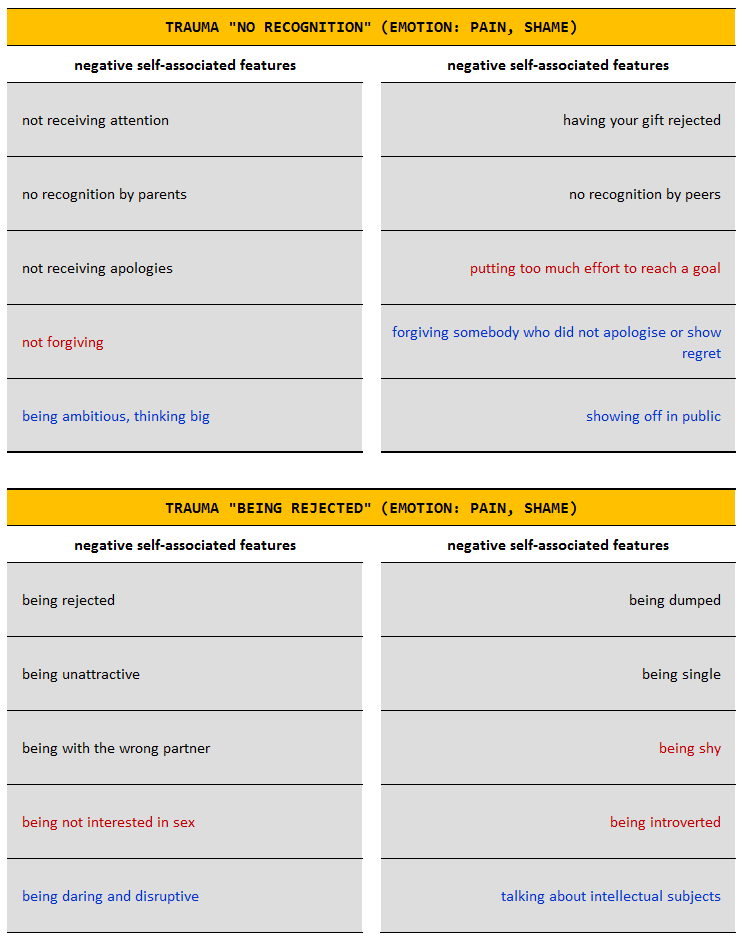}}}
\caption{Some features related to traumas ``NO RECOGNITION'' and ``BEING REJECTED''. Features coloured in black belong to the first category (independent from self), features coloured in red to the second (linked to current self), features coloured in blue to the third (linked to future self).}
\label{norecogn}
\end{center} \end{figure}

\begin{figure}[p] \begin{center} \hspace*{-0.25cm}
{\fboxrule=0.0mm\fboxsep=0mm\fbox{\includegraphics[width=17.50cm]{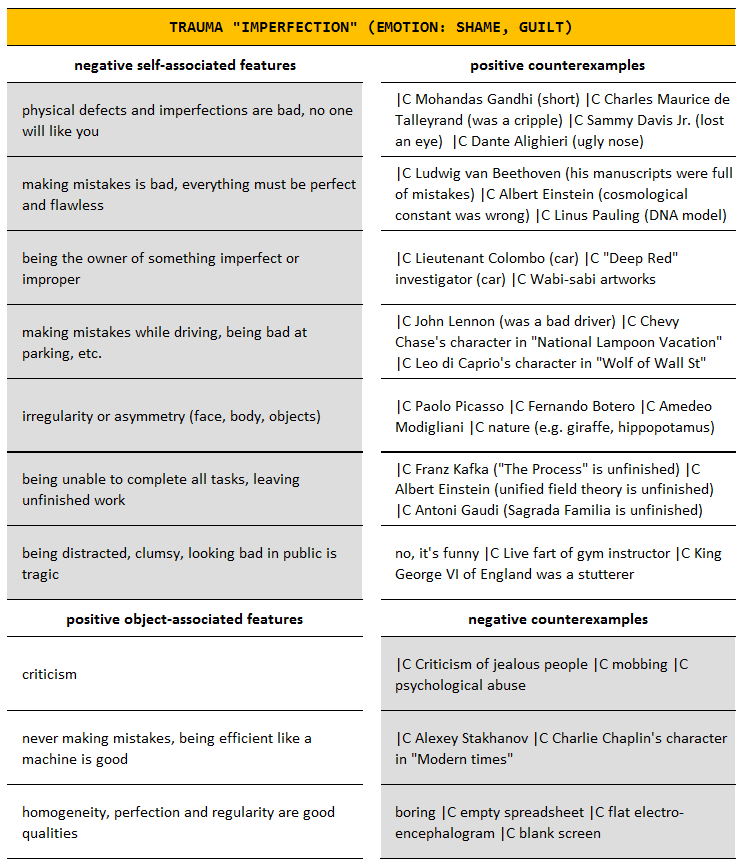}}}
\caption{Counterexamples (symbol ``$\mid$ C'') for features linked to trauma ``IMPERFECTION''. The information reported is correct to the best of the authour's knowledge, does not refer to any specific ``self'' or ``object'', is intended for scientific use only and does not intend to offend any person or entity. As a further precaution, only historical figures and fictional characters are used as examples.}
\label{imperfect}
\end{center} \end{figure}

\begin{figure}[p] \begin{center} \hspace*{-0.25cm}
{\fboxrule=0.0mm\fboxsep=0mm\fbox{\includegraphics[width=17.50cm]{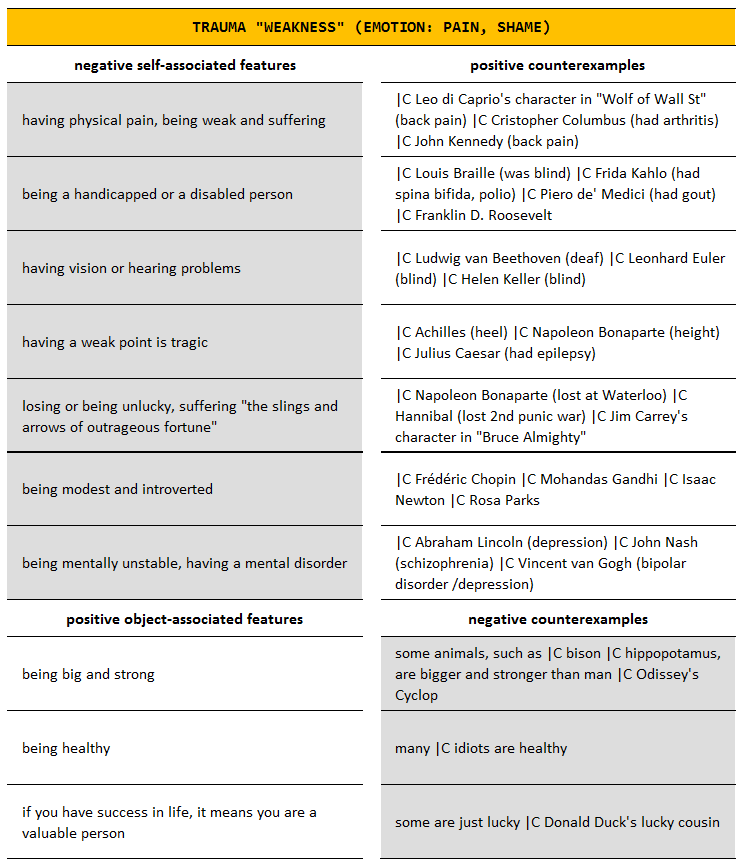}}}
\caption{Counterexamples (symbol ``$\mid$ C'') for features linked to trauma ``WEAKNESS''. The information reported is correct to the best of the authour's knowledge, does not refer to any specific ``self'' or ``object'', is intended for scientific use only and does not intend to offend any person or entity. As a further precaution, only historical figures and fictional characters are used as examples.}
\label{weakness}
\end{center} \end{figure}

\begin{figure}[p] \begin{center} \hspace*{-0.25cm}
{\fboxrule=0.0mm\fboxsep=0mm\fbox{\includegraphics[width=17.50cm]{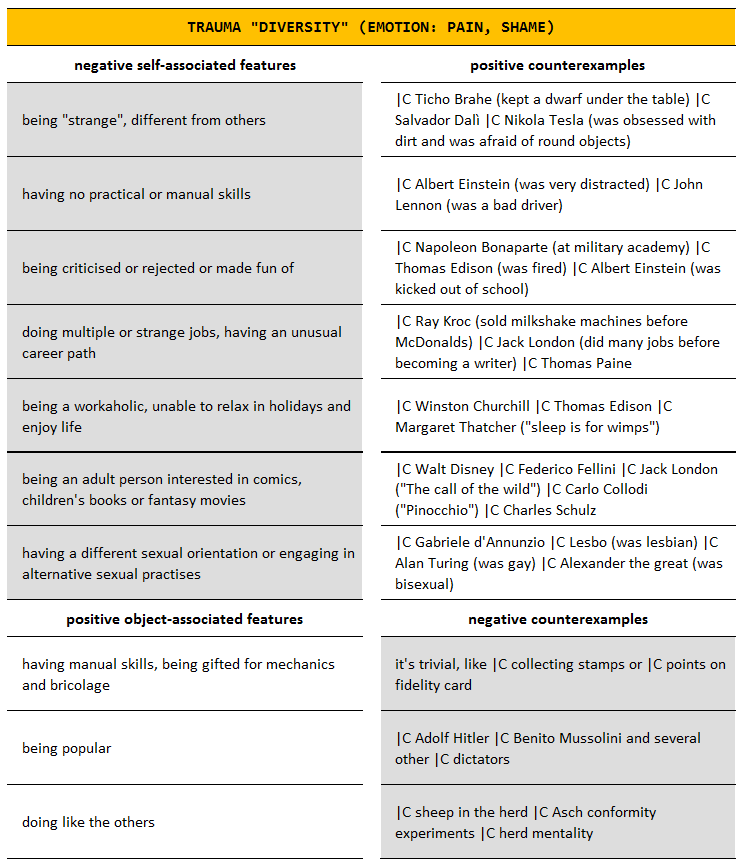}}}
\caption{Counterexamples (symbol ``$\mid$ C'') for features linked to trauma ``DIVERSITY''. The information reported is correct to the best of the authour's knowledge, does not refer to any specific ``self'' or ``object'', is intended for scientific use only and does not intend to offend any person or entity. As a further precaution, only historical figures and fictional characters are used as examples.}
\label{diversity}
\end{center} \end{figure}

\begin{figure}[p] \begin{center} \hspace*{-0.25cm}
{\fboxrule=0.0mm\fboxsep=0mm\fbox{\includegraphics[width=17.50cm]{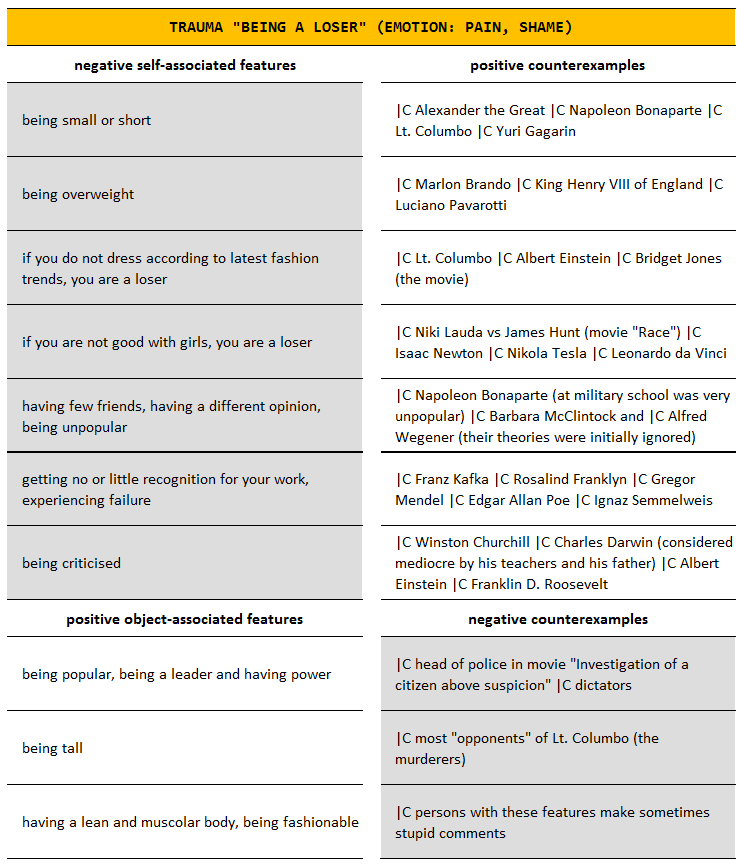}}}
\caption{Counterexamples (symbol ``$\mid$ C'') for features linked to trauma ``BEING A LOSER''. The information reported is correct to the best of the authour's knowledge, does not refer to any specific ``self'' or ``object'', is intended for scientific use only and does not intend to offend any person or entity. As a further precaution, only historical figures and fictional characters are used as examples.}
\label{loser}
\end{center} \end{figure}

\clearpage

\section{Conclusions}

The objective of this work was to propose a model of psychopathology, inspired by Artificial Life and Articial Intelligence. The model is based on two pillars: trauma, which represents the attack to the mind, and dissociation, which represents the defence of the mind. Trauma is always of environmental origin, dissociation is genetically programmed, but is subject to damages, which can be caused by genetic or environmental factors. The insight gained from the model was used to provide an interpretation for various mental conditions: PTSD, complex PTSD, DID, schizophrenia and bipolar disorder. Finally, we have sketched a new therapeutic scheme for psychological trauma, that is expected to be effective for all conditions in which trauma plays a role. A summary of the interpretation is provided in the table below. Further work will be aimed at extending the proposed scheme to other mental conditions.

\vspace*{+0.0cm}
\renewcommand{\arraystretch}{1.2}
\begin{longtable}{m{7.4cm} m{0.0cm} m{7.4cm}} 

\multicolumn{3}{c}{\textbf{PTSD}}\\
\textbf{Traumas.} The traumatic situation is characterised by few very negative features negative. Few contexts are affected (most contexts are unaffected). \newline & &
\textbf{Dissociation.} Dissociation works normally (no genetic or neurodevelopmental issues). Selective dissociation is deployed in traumatic contexts, but is unable to fully patch up the trauma. \\ 

\multicolumn{3}{c}{\textbf{Complex PTSD}}\\
\textbf{Traumas.} Since the traumas occurred early in life and lasted for a long period of time, the traumatic situation is characterised by many negative features. Many contexts are affected. \newline \newline & &
\textbf{Dissociation.} Dissociation works normally (no genetic or neurodevelopmental issues). Selective dissociation is deployed in all traumatic contexts, but is overwhelmed by the pervasiveness of traumas. \newline \\ 

\multicolumn{3}{c}{\textbf{Dissociative Identity Disorder}}\\
\textbf{Traumas.} Since the traumas occurred early in life and lasted for a long period of time, the traumatic situation is characterised by many negative features. Many contexts are affected. \newline \newline \newline & &
\textbf{Dissociation.} Dissociation works normally. Unlike in complex PSTD, in DID the selective dissociation patterns have little in common (the subsets of non-dissociated features are almost completely disjoint), hence the continuity of the self is lost: different personalities take control in different contexts. \\ 

\multicolumn{3}{c}{\textbf{Schizophrenia}}\\
\textbf{Traumas.} The traumatic situation is characterised by several traumatic contexts. Traumas accumulate over the course of life and peak during adolescence. & &
\textbf{Dissociation.} Dissociation is defective (insufficient synaptic connections). Fewer selective dissociation patterns are availbale, dissociation is either too weak or too strong. \\ 

\multicolumn{3}{c}{\textbf{Bipolar Disorder}}\\
\textbf{Traumas.} The traumatic situation is characterised by several traumatic contexts. The statistical distribution of feature values is bipolar. & &
\textbf{Dissociation.} Dissociation is defective (insufficient synaptic connections) and is unable to provide the needed amount of customised dissociation patterns. \\ 

\label{xxxx}
\end{longtable}  

\bibliographystyle{apalike}
\bibliography{lmodapsyc} 

\end{document}